\documentclass[acmlarge]{acmart}

\makeatletter
\newcommand{\myconfshort}{\acmConference@shortname}
\newcommand{\myconffull}{\acmConference@name}
\newcommand{\myconfdate}{\acmConference@date}
\newcommand{\myconfloc}{\acmConference@venue}
\AtBeginDocument{
  \fancypagestyle{firstpagestyle}{
    \fancyhead{}%
    \fancyfoot[C]{}%
  }
  \fancyhf{}
  \fancyhead[LO]{\@headfootfont\shorttitle}%
  \fancyhead[RE]{\@headfootfont\@shortauthors}%
  \fancyhead[LE]{\@headfootfont\footnotesize \myconfshort, \myconfdate, \myconfloc}%
  \fancyhead[RO]{\@headfootfont\footnotesize \myconfshort, \myconfdate, \myconfloc}%
  \fancyfoot[C]{}%
}
\makeatother
\acmBooktitle{\conffull\@ (\confshort), \confdate, \confloc}

\usepackage{color}
\usepackage[normalem]{ulem}
\usepackage{tabularray}
\usepackage{array}
\usepackage{longtable}
\usepackage{caption,subcaption}
\usepackage{soul}

\PassOptionsToPackage{numbers, compress}{natbib}

\AtBeginDocument{%
  }

\copyrightyear{2026}
\acmYear{2026}
\setcopyright{cc}
\setcctype{by-nc-nd}
\acmConference[FAccT '26]{The 2026 ACM Conference on Fairness, Accountability, and Transparency}{June 25--28, 2026}{Montreal, QC, Canada}
\acmBooktitle{The 2026 ACM Conference on Fairness, Accountability, and Transparency (FAccT '26), June 25--28, 2026, Montreal, QC, Canada}
\acmDOI{10.1145/3805689.3812411}
\acmISBN{979-8-4007-2596-8/2026/06}

\begin{document}

\title[Evaluating Structured Documentation as a Tool for Reflexivity]{Evaluating Structured Documentation as a Tool for Reflexivity in Dataset Development}

\author{Eshta Bhardwaj}
\authornote{Both authors contributed equally to this research.}
\email{eshta.bhardwaj@mail.utoronto.ca}
\orcid{0000-0001-8523-5201}
\affiliation{%
  \institution{University of Toronto}
  \streetaddress{}
  \city{Toronto}
  \state{Ontario}
  \country{Canada}
  \postcode{}
}
\author{Ciara Zogheib}
\authornotemark[1]
\email{ciara.zogheib@mail.utoronto.ca}
\orcid{0000-0002-0620-3007}
\affiliation{%
  \institution{University of Toronto}
  \streetaddress{}
  \city{Toronto}
  \state{Ontario}
  \country{Canada}
  \postcode{}
}

\author{Christoph Becker}
\email{christoph.becker@utoronto.ca}
\orcid{0000-0002-8364-0593}
\affiliation{%
  \institution{University of Toronto}
  \streetaddress{}
  \city{Toronto}
  \state{Ontario}
  \country{Canada}
  \postcode{}
}

\renewcommand{\shortauthors}{Bhardwaj and Zogheib, et al.}

\begin{abstract}
It is prominently recognized that dataset development in machine learning is a value-laden process from problem formulation to data processing, use, and reuse. Structured documentation frameworks such as datasheets, data statements, and dataset nutrition labels have been created to aid developers in documenting how their datasets were produced and, according to the creators of the frameworks, to facilitate reflexivity in dataset development. While reflexivity is a stated goal, it is unclear whether and to what extent these structured dataset documentation frameworks incorporate concepts from reflexivity literature (at FAccT and elsewhere) and whether the use of the frameworks demonstrates reflexivity. Here, we adopt mixed-method thematic analysis and corpus-assisted discourse analysis to explore how reflexivity is incorporated in structured documentation frameworks and their responses. We demonstrate empirically that there is a general lack of engagement with major themes of reflexivity in both dataset documentation frameworks and published applications of these frameworks. We present a codebook of major reflexivity topics, recommend actionable strategies, and propose a set of extended datasheet questions to more effectively incorporate these topics into structured documentation frameworks and in the FAccT literature.
\end{abstract}



\maketitle

\section{Introduction} \label{introduction}

It is well-established that dataset development in machine learning is a value-laden process. From problem formulation to data processing, each step consists of decisions that have real-world effects and far-reaching impacts \cite{croskey_liberatory_2025, li_our_2025, poirier_accountable_2022, poirier_what_2025, rifat_data_2024, Scheuerman_Hanna_Denton_2021, simson_lazy_2024}. Because of this impact, there have been calls for increased reflexivity in the dataset development process through mechanisms of documentation \cite{miceli_documenting_2021}. Miceli et al. recognize the process of documentation as reflexivity in two ways: the documentation produced “eases communication and promotes organizational accountability” and “the act of documenting [is] intended to make naturalized preconceptions and routines explicit” \citep[p.~168]{miceli_documenting_2021}. Several structured documentation frameworks have been developed to support dataset creators in carrying out this reflection.  

While structured documentation frameworks expect dataset developers to reflect on data design decisions, it is unclear whether and to what extent these frameworks encompass key concepts from the broader \textit{reflexivity} literature and whether the frameworks contribute to practices of reflexivity as it is understood in that literature. Given the widespread adoption of datasheets as the standard for dataset documentation and the framework’s explicit goal to increase developers’ reflection, this is an important juncture at which to assess whether current structured documentation frameworks are effectively prompting reflexive data practices. 

We explore whether and how structured documentation frameworks demonstrate reflexivity. We additionally analyze responses to documentation templates to understand the current standard for reflexivity within dataset development in machine learning (ML). Our goal is to ascertain whether reflexivity is possible through structured documentation and to what extent it is currently being practiced by dataset developers. Our research questions allow us to explore the topic along the theory-practice continuum:  

\begin{enumerate}
    \item How is reflexivity conceptualized within the FAccT literature? (Theory)
    \item How are the identified themes illustrated within structured documentation frameworks? (Theory-informed guidance for practice)
    \item How do dataset developers demonstrate reflexivity through responses to documentation frameworks? (Practice)
\end{enumerate}

Our analysis is conducted in three stages. We first perform reflexive thematic analysis (RTA) on a corpus of 30 documents (books, journal articles, and conference proceedings) to identify key themes that reoccur within reflexivity literature regarding how reflexivity is conceptualized. In the second stage, we apply to RTA to search for these themes within three structured dataset documentation frameworks (datasheet \cite{gebru_datasheets_2021}, data statement \cite{mcmillan-major_data_2024}, and dataset nutrition label \cite{chmielinski_dataset_2022}). We lastly perform corpus-assisted discourse analysis on completed documentation templates (i.e., datasheets of published datasets) to assess whether the themes are considered by dataset developers when documenting datasets. 

We develop a codebook of six themes from our thematic analysis and synthesize reflexivity literature across various domains of discourse. Our analysis of the occurrence of these themes in structured documentation templates reveals that while \textit{reflexivity} is explicitly stated as a goal, the questions largely prompt dataset creators for \textit{transparency} about their dataset development choices. Similarly responses are transparent about dataset creation and identify where potential biases and negative impacts could occur but do not address reflexivity.

\textbf{Contributions: }By synthesizing literature on reflexivity, we broaden the conceptualization of reflexivity at FAccT (RQ1). We analyze structured documentation frameworks through the lens of reflexivity first, thus we are able to identify the priorities for documentation implicitly embedded in structured documentation frameworks (RQ2). By analyzing responses, we identify aspects of reflexivity currently not present within practices (RQ3). Our paper thus summarizes ways in which structured documentation facilitates reflexivity and where it falls short. We  provide recommendations for how datasheets, as a popular example of structured documentation, can better prompt dataset developers to reflect on their dataset development processes.

\section{Background} \label{background}

We first introduce how reflexivity is discussed and conceptualized in feminist and poststructuralist works, critical data studies, broader methods literature, and others, and then look at how research examining dataset documentation practices have begun advocating for the inclusion of reflexivity. 

\subsection{Understanding Reflexivity} \label{background_reflexivity}

Defining reflexivity is an ongoing scholarly project, but existing definitions tend to coalesce around similar concepts: England tells us that reflexivity is “self-critical sympathetic introspection and the self-conscious analytical scrutiny of the self as researcher” \citep[p.~82]{england_getting_1994}, an idea echoed nearly thirty years later by Jamieson et al., who define reflexivity as “the process of engaging in self-reflection about who we are as researchers, how our subjectivities and biases guide and inform the research process, and how our worldview is shaped by the research we do and vice versa” \citep[p.~2]{jamieson_reflexivity_2023}. 

Discussions of reflexivity have their roots in qualitative scholarship, with the largest part of the discourse situated in the context of poststructuralist and constructivist thought \cite{kenway_bourdieus_2004}. In this vein, scholars including Harding \cite{harding_after_1992}, Haraway \cite{haraway_situated_1988}, Pillow \cite{pillow_confession_2003}, and Day \cite{day_reflexive_2012} have applied feminist and critical lenses to articulate how the professed value-neutrality and ahistorical nature of science merely normalizes the priorities or standards of dominant social and political groups, often at the expense of situated or marginalized forms of knowledge. Among the major theorists in this area, Bourdieu \cite{bourdieu_science_2004} offers a marked contrast: rather than engaging with feminist and poststructuralist reflexivity, Bourdieu focuses on the scientific field as a disciplining and corrective force on individual scholars and takes the stance that reflexivity, rather than dismantling ideas of value-neutrality, serves to “buttress the epistemological security” \citep[p.~528]{kenway_bourdieus_2004} of science’s ability to produce trans-historical truths.

Comparatively recently, there has been movement towards reflexivity in data-centric and technical research contexts \cite{boyd_quinta_2023, cambo_model_2022, tanweer_why_2021,Allhutter_Berendt_2020}. In translating reflexive practices into these settings, this work confronts new issues: how can “scrutiny of the self as researcher” be applied in fields where stances of objectivity and scientific distance are still the dominant conventions \cite{cambo_model_2022}? How do well-established discourses about transparency in data sciences fit into reflexive practice – and are the two concepts, as Boyd argues, as interchangeable as they are often implied to be \cite{boyd_quinta_2023}? Broadly speaking, reflexivity in data-centric and technical research serves as a challenge to perceptions of data work as objective and insulated from social, historical, and cultural context.

A look at FAccT publications provides insights as to how reflexive data work is engaging (or, as it were, not engaging) with broader discourses of reflexivity. Of the 84 FAccT papers (listed in Appendix \ref{appendixa}) that mention reflexivity across all archived years, a handful cite the constructivist and feminist reflexivity literature from scholars like Haraway, but a much larger portion cite Bourdieu’s reflexivity. Benbouzid goes so far as to suggest that computer science as a whole specifically considers reflexivity as conceptualized by Bourdieu \cite{benbouzid_fairness_2023}. This trend is in direct contrast to wider discourses on reflexivity \cite{kenway_bourdieus_2004}, and raises the question: if scholarship on data and data practice is engaging selectively with one portion of the literature on reflexivity, how does this impact the way that reflexivity in data work is put into practice?

\subsection{Dataset Documentation Practices} \label{background_documentation}

Machine learning research has increasingly turned towards the improvement of data to bolster model results and fundamental understanding. Research has accordingly become more focused on data-centric machine learning and the construction of datasets (e.g., \cite{dhamala_bold_2021, jourdan_fairtranslate_2025, li_our_2025}). Despite the significance of these data processes, challenges remain around the lack of accountability and transparency in how models are developed and deployed as well as their outcomes \cite{Diakopoulos_2016, Hutchinson_2021, Khan_Hanna_2022, Raji_2020, Veale_Van_2018}. To address this, research proposing frameworks to aid in the documentation of datasets’ contents and data design decisions has become prominent. 

These frameworks often take the form of context documents (“interventions designed to accompany a dataset or ML model, allowing builders to communicate with users” \citep[p.~2]{boyd_datasheets_2021}. They are used to provide documentation for datasets by detailing aspects of provenance and data collection and are particularly geared to answering ethical questions about the data. The most popular of these structured documentation frameworks is datasheets, first introduced in 2018 \cite{Gebru_2018} and revised in 2021 \cite{gebru_datasheets_2021}. Datasheets provide a method for researchers to be intentional in their process through the documentation of dataset context and contents, thus increasing transparency and accountability \cite{gebru_datasheets_2021}. In the 7 years since its release, datasheets have become a standard within industry, government, and academia (e.g., the NeurIPS Datasets and Benchmarks track between 2021-2024 recommended datasheets to document dataset contents and uses, leading to the development of hundreds of datasheets \cite{noauthor_call_2021,noauthor_call_2022,noauthor_call_2023,noauthor_call_2024}). It has led to the creation of new documentation frameworks developed including data statements \cite{mcmillan-major_data_2024}, data cards \cite{pushkarna_data_2022}, and dataset nutrition labels \cite{chmielinski_dataset_2022}, context-specific datasheets (e.g., \cite{papakyriakopoulos_augmented_2023, Connolly_Hueholt_Burt_2025, rostamzadeh_healthsheet_2022}), as well as numerous studies examining the usage and application of datasheets (e.g., \cite{bhardwaj_state_2024,boyd_datasheets_2021,schramowski_can_2022}). 

It is widely recognized that the process of dataset development reveals values \cite{Scheuerman_Hanna_Denton_2021}, requires tacit knowledge and data skills \cite{Hutchinson_2021, muller_forgetting_2022, muller_how_2019}, is embedded with nuance and craft work \cite{jo_lessons_2020, thomer_craft_2022}, and consists of decision-making at every step. Dataset development is a highly subjective process and the outcomes from using a given dataset reveals those choices \cite{gonzalez_zelay_towards_2019, Passi_Barocas_2019}. Accordingly, developers should engage with reflexive practices to more critically consider how their personal, social, historical, cultural, and intellectual influences impact the dataset development process. Being reflexive situates dataset development in context and can improve the quality of datasets by making the tacit explicit and by reflecting on bias and power dynamics. Documenting this reflexivity then further promotes rigorous data work and showcases how reflexive data work can be done practically. 

The datasheets framework explicitly states promoting reflexivity as an intended objective: “For dataset creators, the primary objective is to encourage careful reflection on the process of creating, distributing, and maintaining a dataset, including any underlying assumptions, potential risks or harms, and implications of use.” \citep[p.~86-87]{Gebru_2018}. Thus we examine whether the datasheet template prompts creators to consider reflexivity and whether and how dataset creators engage in and document their reflexive data practices.

\section{Methods} \label{methods}

\begin{figure*}[h!]
    \centering
    \includegraphics[width=1\linewidth]{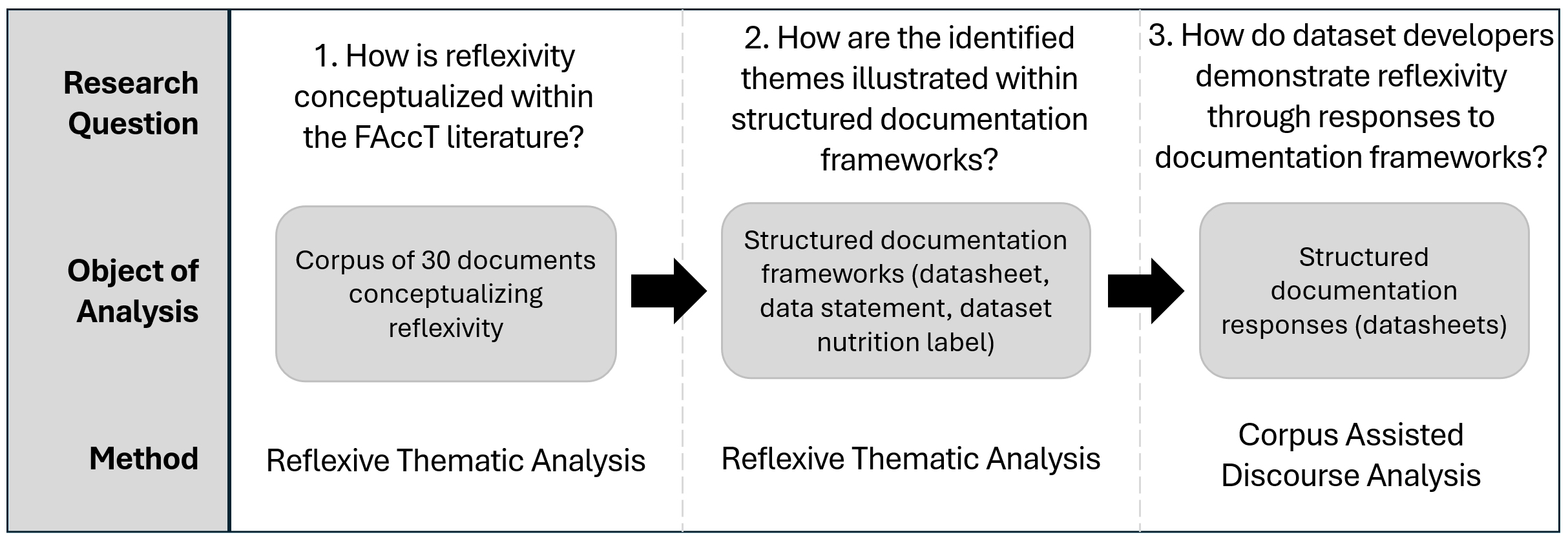}
    \caption{Our multi-stage approach to conceptualizing reflexivity and analyzing it within structured documentation frameworks and dataset documentation.}
    \label{fig:fig1}
\end{figure*}

We take a mixed-methods approach to exploring reflexivity in dataset documentation along the theory-practice continuum: reflexive thematic analysis (RTA) of the literature on reflexivity to understand how reflexivity is conceived in theory (RQ1), followed by RTA of structured dataset documentation templates (RQ2), and lastly corpus assisted discourse analysis (CADA) of filled-out examples (RQ3). Our approach is summarized in Figure \ref{fig:fig1}. 

\subsection{Reflexivity in the Literature} \label{methods_reflexlit}

The first stage of our study addresses RQ1 by analyzing a corpus of reflexivity literature (listed in Appendix \ref{appendixa}) to identify the key themes through which reflexivity is conceptualized. We comprised this corpus by applying an interpretive synthesis literature review approach \cite{dixon-woods_conducting_2006, flemming_synthesis_2010, mcferran_critical_2016, xie_meta-synthesis_2020}. Interpretive synthesis reviews are optimal in cases where the topic of study – in this case, reflexivity – is defined and discussed disparately across domains and bodies of literature. In these cases, conventional keyword-searching systematic review approaches can overlook relevant discussions and publications. Interpretive synthesis methods allow for ‘big picture’ overviews of a topic, incorporating researcher expertise and facilitating resolution and comparison across “competing schools of thought” \citep[p.~15]{allard_little_2024}, as we observe in the FAccT literature’s engagement with Bourdieu reflexivity and post-structuralist reflexivity more broadly.

Adopting literature search strategies from \cite{allard_little_2024, dixon-woods_conducting_2006, Desrochers_2017}, we compiled the corpus starting with reviewing mentions of the word “reflexiv-” (including reflexivity and reflexive) in all FAccT conference proceedings. We included those papers that had meaningful contributions to conceptualizing reflexivity in the corpus. We used ‘pearl-growing’ and citation mining techniques to identify works cited by FAccT authors when discussing or introducing the concept of reflexivity. We  added purposefully to this corpus by including related discourse on reflexivity including publications from feminist reflexivity, critical data studies, and the broader methods literature.

We applied a RTA approach to identify recurring topics in our corpus of literature (n = 30) following the methods of Braun and Clarke \cite{braun_can_2021, braun_reflecting_2019, byrne_worked_2022}. This work took place iteratively. We started the process by reading, discussing, and coding one paper synchronously and collaboratively to suggest initial codes and code definitions. Then two of the authors each reviewed half of the documents and independently performed open coding. During and after coding, we discussed and deliberated on codes as we consolidated them into themes; rather than calculating intercoder reliability, we met to discuss any inconsistencies and reach a consensus, following O’Connor and Joff's procedures for qualitative research \cite{o2020intercoder}.

\subsection{Reflexivity in Structured Dataset Documentation Templates} \label{methods_reflexstrucdoc}

In the second stage, we address RQ2 by analyzing whether the key themes conceptualizing reflexivity (derived from the previous stage) are exhibited within three structured documentation frameworks (including the templates they introduce and their published papers) by applying RTA \cite{braun_can_2021, braun_reflecting_2019, byrne_worked_2022}. We specifically chose to analyze datasheets \cite{gebru_datasheets_2021}, data statements \cite{mcmillan-major_data_2024}, and dataset nutrition labels \cite{chmielinski_dataset_2022} as these are the most commonly expected documentation formats within the NeurIPS Datasets and Benchmarks (D\&B) track \cite{noauthor_call_2021,noauthor_call_2022,noauthor_call_2023,noauthor_call_2024, mcmillan-major_data_2024}. 

Two authors independently reviewed each of the 3 frameworks and performed open coding by highlighting those components that reflected the themes gleaned from stage 1. The codes were discussed collaboratively to synthesize how themes are demonstrated within each of the documentation frameworks.

\subsection{Reflexivity in Dataset Documentation at NeurIPS} \label{methods_reflexneurips}

To analyze whether scholars are using structured data documentation frameworks as tools for reflexivity, and how they are doing so, we conducted corpus-assisted discourse analysis (CADA). CADA is a framework combining quantitative and qualitative text analysis methods to provide a holistic understanding of language as it is being used in particular contexts \cite{georgakopoulou_corpusassisted_2020}. The benefit of CADA is in its ability to facilitate both broad, corpus-level analysis of trends in texts, and ‘closer’, context-rich readings to explain those trends \cite{liu_climate_2022}. 

We make use of a collection of papers published in the NeurIPS Datasets and Benchmarks (D\&B) track from 2021-2023\, listed according to number in Appendix \ref{appendixb} and referred to as such in the remainder of the paper. We use NeurIPS datasheets as a data source because of the quantity of datasheets, NeurIPS’ position as a top ML venue with a dataset track, and because this dataset has previously been used for dataset documentation research by Bhardwaj et al. \cite{bhardwaj_state_2024}. The corpus analyzed here consists of those papers which include completed structured documentation (n = 36, of which 35 are datasheets and one is a data statement). We conduct standard pre-processing of the text including removal of stopwords, lemmatization, and filtering of non-alphabetic terms (e.g., numbers and hyperlinks).

Quantitative text analysis includes the calculation of descriptive metrics such as term frequency (which words occur most often), collocations (which pairs and triplets of words co-occur most often), and lexical diversity (ratio of unique words to total words). We also conduct topic modeling using latent dirichlet allocation (LDA) and calculate semantic similarity to compare across datasheets (i.e., for datasheet 1 vs datasheet 2), and across questions (i.e., for Q1 across all filled-out datasheets vs. Q2 across all filled-out datasheets), following guidance for topic modeling from \cite{gan_selection_2021, mattingly_introduction_2021}. Quantitative text analysis is completed in Python using libraries NLTK \cite{bird_natural_2009}, gensim \cite{Sojka_2010}, and Scikit-learn \cite{Pedregosa_2011}. 

We complement these quantitative analyses using qualitative, in-depth readings of the filled-out structured data documents. Qualitative engagement with these texts included exploration of themes previously identified during RTA and identification of particularly noteworthy illustrative examples of strong or weak reflexivity. We adopt the interpretive synthesis strategies described by Noblit and Hare \cite{Noblit_Hare_1988} and Dixon-Woods et al. \cite{dixon-woods_conducting_2006} to integrate qualitative and quantitative findings by identifying contradictions, shared conclusions, and explanations.

\section{Findings} \label{findings}

We present our findings in three sections. We present the codebook of the themes elicited from reflexivity literature, discuss that structured documentation frameworks currently do not support or prompt dataset creators to be reflexive, and accordingly find that the responses do not demonstrate reflexivity but rather transparency. 

\subsection{Emergent Themes of Reflexivity} \label{findings_themes}

Our thematic analysis of reflexivity literature yielded six themes encompassing topics that recur as key features of its conceptualization. The themes are not mutually exclusive and frequently relate to each other in the reviewed texts. For example, a paper describing how reflexive research pays attention to social, cultural, or historical context is also likely to discuss how that context shapes power dynamics between researchers and research participants. We define and present these themes in Table \ref{tbl:tbl1}. The extended codebook with examples is provided in Appendix \ref{appendixc}.

\NewTblrTheme{mytheme}{
  \SetTblrTemplate{capcont}{empty}
}

\definecolor{Black}{rgb}{0,0,0}
\begin{longtblr}[
  theme = mytheme,
  caption = {Themes elicited through RTA that summarize the conceptualization of reflexivity within the compiled corpus.},
  label = {tbl:tbl1},
]{
  width = \linewidth,
  colspec = {Q[190]Q[750]},
  hlines,
  vlines = {Black},
  vline{1} = {-}{black},
}
Theme & Description\\
Marginalization and power & This theme disentangles the roles assigned within knowledge production and how their placements within social and/or scholastic hierarchies result in unequal impacts.~\\
Attention to historical context & This theme positions research as situated historically, conducted in a particular time and place with accompanying social, political, and cultural context. Reflexive research does not obscure or deny this historicity.\\
Attention to scholarly context & This theme emphasizes how academic disciplines and research domains constitute fields with norms, incentives, and internal epistemological and evidentiary standards (all of which can shape scholarly work).~\\
Questioning claims of neutrality & This theme rejects the notion of neutrality or objectivity within research and science. Claims of neutrality are opposed with embracing situatedness and claiming it a stronger form of objectivity, i.e., reflexivity makes research more objective and thus more legitimate by acknowledging situatedness.~\\
Moving beyond individual positionality to consider relationality & This theme prompts for considering positionality in more active ways and reflexivity beyond positionality.~\\
Accountability beyond transparency & This theme differentiates between accountability and transparency as features of reflexive research where accountability is taking ownership of the research and its potential negative repercussions while transparency is considered the mere identification of biases and harmful impacts.
\end{longtblr}

We elaborate on how each theme is illustrated in the literature and perspectives that informed the themes.

\textbf{Marginalization and power.} Discussions of this theme in the literature state that reflexivity should consider the multiple hierarchies of power (between researcher and researched and between researcher groups), the consequence of hierarchical power dynamics that value certain groups’ knowledge as valuable or prioritize pursuing questions of importance to certain groups, the inequality of access to knowledge production processes, the tokenism and appropriation of marginalized points of view, and the treatment of those researched as one homogenous entity. 

\textbf{Attention to historical context.} In the literature, this theme took the form of arguments that reflexive work is that which is intentionally situated in a particular time, place, and socio-cultural setting. ‘Attention to historical context’ includes acknowledgement of “crude aspects” of researcher identity (e.g., race, nationality, class) \citep[p.~200]{mccorkel_what_2003}, but also of the motivation for and impacts of the research in society more broadly \cite{jamieson_reflexivity_2023}.   

\textbf{Attention to scholarly context.} This theme expresses the commitment in the reflexivity literature to engagement with the domain and institutional contexts in which research is carried out. It includes discussions on how the act of approaching research from an ‘analytic disposition’ more broadly shapes problem identification and formulation.  Reflexive work, it is argued, is critical of how the act of situating scholarship limits how questions are asked and evaluated in a given domain \cite{hirsbrunner_critical_2024}. This theme is very prominent in Bourdieu-inspired discussions of reflexivity \cite{hess_neoliberalism_2013}. 

\textbf{Questioning claims of neutrality.} Reflexivity literature 1) critiques knowledge production processes while advocating for them to generate less distorted analyses, 2) surfaces tensions in how select types of research are prioritized by dominant groups, and 3) argues that research from the standpoints of the marginalized is powerful to prevent “views from nowhere” \cite{haraway_situated_1988}. This theme also encompasses viewpoints that state that reflexivity is not a metric for legitimate or rigorous science. For example, scholars argue that lived experience discussed through reflexivity or positionality statements does not automatically result in “high quality” research \cite{schroeder_disclosure_2025}. 

\textbf{Moving beyond individual positionality to consider relationality.} Some of the discussions have raised concerns that reflecting on positionality as a form of post-hoc self-disclosure of demographics is limited in its impact and could be considered performative (as compared to active reflexivity which is fluid, dynamic, continuous and also questions how others perceive their positionality). The literature also points to challenges of self-reflexivity including how it places vulnerable researchers at risk, may result in biases during peer review, and expects researchers to somehow “make conscious” unconscious biases during research. 

\textbf{Accountability beyond transparency.} Transparency is discussed as a component of reflexivity in the broader literature \cite{seaver_vertesi_knowing_2019} and the literature on reflexivity in data work \cite{DIgnazio_Klein_full2020}. This theme captures the common conclusion of these discussions: that transparency, as a documentation of process, is not in itself a reflexive practice and thus insufficient, requiring accountability as well. 

\subsection{Themes in Structured Documentation Templates} \label{findings_themesstrucdoc}

After identifying key themes that characterize reflexive practice, we qualitatively analyzed the occurrence of these themes in three widely used structured dataset documentation frameworks (including the documentation templates and the publications describing their creation): datasheets, data statements, and dataset nutrition labels.

\subsubsection{Reflexivity as Transparency}\label{findings_transparency}

First, to the broader relevance of reflexivity in these documentation frameworks: the creators of datasheets explicitly state that a “primary objective” of datasheets is to “encourage careful reflection on the process of creating, distributing, and maintaining a dataset”, in dataset creators \citep[p.~86]{gebru_datasheets_2021}, and “even alter this process in response to their reflection” \citep[p.~88]{gebru_datasheets_2021}. This very much situates datasheets as a tool for reflexivity. Both data statements and dataset nutrition labels take inspiration from and repurpose questions from datasheets \cite{chmielinski_dataset_2022, mcmillan-major_data_2024}; we can conclude that reflexivity is a goal of these frameworks, and that we might accordingly expect to find our reflexivity themes incorporated throughout. 

The reality is somewhat different. Of the themes from the reflexivity literature, the one most meaningfully absent across all structured dataset documentation templates is ‘accountability beyond transparency’. All three frameworks (datasheets, dataset nutrition labels, and data statements) list transparency as part of their motivation, and by providing the means to document procedural decisions and details, structured documentation frameworks do support more transparent data work. The reflexivity literature, however, while embracing transparency and accountability as key components of reflexivity, makes clear that they are not, without engagement with the other substantive features of reflexivity, sufficient to make work reflexive. This broader engagement with reflexivity is absent from the structured dataset documentation tools analyzed here. To the extent that any of the frameworks reflect on accountability beyond transparency, they do so in terms of how structured documentation frameworks “provide an opportunity for dataset creators to distinguish themselves as prioritizing transparency and accountability” \citep[p.~92]{gebru_datasheets_2021}, or of how completion of these documents allows dataset creators to meet mandates of conferences or publication venues \cite{mcmillan-major_data_2024}. Transparency without meaningful reflection that leads to changes in practice or lessens harmful impacts does not in itself fulfill expectations for reflexivity. 

\subsubsection{Reflexivity without Reflection?} \label{findings_reflection}

We can observe some similar trends in how each of the three data documentation frameworks engages with different reflexivity themes. The dataset nutrition labels arguably demonstrate the ‘questioning claims of neutrality’ theme by criticizing the “assumed neutrality” \citep[p.~1]{chmielinski_dataset_2022} of data and technical work and including questions about the use of proxy characteristics or involvement of communities represented in the dataset; likewise, data statements, perhaps due to their focus on text data, repeatedly devote attention to historical context. Occurrences like these, of particular themes in different documentation frameworks, share the same underlying issue: where frameworks do touch on reflexivity themes, they do so by asking dataset creators to list positions or decisions, but do not go beyond this to prompt creators to consider implications.

In effect, what we observe in datasheets, data statements, and dataset nutrition labels is the opposite of the reflexive practice captured in the theme ‘moving beyond individual positionality to consider relationality’. Pillow critiques attempts at reflexivity that “result in a simple identifying of oneself or a telling of a confessional tale” \citep[p.~182]{pillow_confession_2003}; this listing of individual positions and decisions without deeper reflection is what is prompted by the structured documentation templates. For example, section 8 of the data statements framework, with the aim of showing how “the time, place, and cultural context allow for deeper understanding of how the texts collected relate to their historical moment” \cite{mcmillan-major_data_2024}, prompts dataset creators to list place, date, and ‘non-linguistic context’ (reflecting the ‘awareness of historical context’ theme, at least on the surface), but does not invite further reflection on how any of these listed positions might have impacted the resulting textual data. Alternately, datasheets’ Q24 addresses who was involved in data collection and how they were compensated. This topic is very relevant to the theme of ‘marginalization and power’ that reoccurs throughout the reflexivity literature, and is certainly germane to the field of data work (see scholarship on ongoing issues with compensation and credit for data work \cite{davies_care_2024, muller_forgetting_2022}), but the question as it exists does not prompt dataset creators to reflect on the implications of their answer with regards to how compensation can perpetuate, for example, existing cycles of exploitation and marginalization of data workers. 

Overall, and with attention to RQ2 (‘How are the identified themes illustrated within structured documentation frameworks?’), the papers about the three frameworks confirm reflexivity as an intended function, and questions and prompts in the structured documentation templates touch on some of the themes of the reflexivity literature, but we do not observe that any of the frameworks explicitly prompt the self-critical, relational engagement that characterizes ‘strong’ reflexivity. That is: much of whether structured dataset documentation frameworks will improve reflexivity is left to the initiative of the dataset creators using the frameworks.

\subsection{Themes in Structured Documentation Responses} \label{findings_themesresponses}

\subsubsection{Quantitative Results} \label{findings_quantCADA}

We began our exploration of how dataset creators are or are not using structured data documentation tools for reflexivity quantitatively, by conducting an array of descriptive analyses on a sample of 36 completed datasheets, following \cite{bird_natural_2009}. Keyword searching of terms related to the reflexivity themes previously identified (see Appendix \ref{appendixd} for the full results) demonstrates low usage of reflexivity-specific language. Exploration of term frequency reveals some potential explanations for this: ‘yes’ is one of the most frequently used words across all datasheets, and several datasheets include ‘no’ or ‘not applicable’ answers with no further explanation. Documentation is taking place, because answers are being provided, but one-word answers are unlikely to contain in-depth reflection. For example: in datasheet Q2 and Q3, which are concerned with research funding and ownership, major technology company names are among the most frequently occurring terms. The implications of corporations as primary funders of dataset creation might be productively interrogated during reflexive practice, but is not at present being elaborated on.

The absence of key terms also means that certain quantitative methods common to CADA such as the identification of collocations produce trivial results (see full outputs in Appendix \ref{appendixd}). We apply those quantitative approaches that we can cautiously: across the entire corpus, we observe a lexical diversity, a ratio of unique words to total words, of only 0.16, indicating very low variety in the language being used between datasheets and between questions. Likewise, LDA topic modeling of the corpus resulted in low intertopic distance and coherence scores of less than 0.3, suggesting that the documents being evaluated are strongly homogenous topically.

Full topic modeling results are included in Appendix \ref{appendixd}; for our focus on reflexivity, it is more interesting and informative to interrogate \textit{why} topic modeling results are unreliable. How can structured documents describe completely distinct projects, yet cannot inform meaningfully different topics? Scholars like Schmidt \cite{schmidt2012words} and Gillings and Hardie \cite{gillings2023interpretation} have criticized the stability of topic models for text evaluation more generally; while, as they suggest, preprocessing and unstable clusters may impact outcomes,we observe that outside of topic modeling results, dataset creators’ answers are still dominated by the same recurring answer trends, like the one-word answers observed through term frequency. If we were seeing deep and nuanced reflexivity taking place for different dataset projects, we might expect unique words to be more important in the documentation of these projects, or to see more meaningfully distinct topics of discussion. In the next section, we turn to in-depth, qualitative document analysis to challenge and further explore this inference.

\subsubsection{Qualitative Results} \label{findings_qualCADA}

An in-depth review of datasheet responses (as the one of the three reviewed documentation templates most represented in our corpus) demonstrated that datasheets are exemplified as an exercise in transparency. This is particularly evident for certain questions which prompt creators to discuss potential negative impacts, harmful uses, biases, or errors resulting from specific dataset development choices. These questions relate directly to the theme of ‘accountability beyond transparency’. However, responses to these questions focus largely on transparency by identifying the issues but not reflecting on them. For example, in response to Q6 about conducting ethical review processes, our analysis revealed that documentation often only stated the IRB name and rarely described the review processes despite the question prompting for it specifically (e.g., datasheet 10, 14, 19, 21, etc.). For Q41 about “tasks for which the dataset should not be used”, datasheets commonly documented “no” or “not applicable” with no further reflection (e.g., datasheet 6), stated tasks for which the dataset should not be used but did not explain why (e.g., datasheet 17), or did not provide mitigation strategies (e.g., datasheet 13). 

While the theme of ‘accountability beyond transparency’ was best represented in the datasheet responses (albeit for solely showcasing transparency), all other themes were minimally or not at all represented, supporting quantitative results that pointed to a lack of in-depth reflection. Minimally represented themes included ‘marginalization and power’ which was documented by disclosing wages or compensation paid towards crowdworkers or data annotators (e.g., datasheet 5, 10, 12, 19, etc.) but did not discuss any potential power dynamics or provide additional detail that would contextualize the compensation as fair or not. 

The theme ‘attention to historical context’ was only present in one question of the datasheet – Q26 (“Were any ethical review processes conducted (e.g., by an institutional review board)?)” \cite{gebru_datasheets_2021}. We link this question to the `historical context' theme as well as to `marginalization and power' and `accountability beyond transparency'; in the reflexivity literature, all three are often discussed in terms of harms historically committed against particular communities, and the ethical responsibility of researchers to avoid continuing these harms. Similar to Q26, responses to this question (if the dataset contained data about people) focused solely on identifying the review board and the outcome of the review process but did not demonstrate context awareness.

‘Attention to scholarly context’ was present in responses that discussed the rationale behind data splits (Q12), such as datasheet 4, 5, and 20, and identified specific expertise involved in the data collection or annotation process (Q23 and Q24), such as datasheet 2, 3, and 15. For example, datasheet 18 mentions that volunteer student residents of the local townships were recruited to aid in the annotation process to label images of neighbourhoods. 

The themes ‘questioning claims of neutrality’ and ‘moving beyond individual positionality to consider relationality’ were not strongly represented in any of the responses as no datasheets had individual or collective positionality statements. There was some consideration given to the situatedness and non-neutrality of decisions related to sampling which reflect scholarly context and individual experience and training. Datasheet responses discussed how their sampling strategies were not representative but instead biased. For example, datasheet 32 stated: “We are aware of the fact that biases exist inherently in our data collection, for the following reasons: (a) the data was extracted from a certain population, period and region, with certain manufacturers, (b) the annotations were made by radiologists from a certain region, (c) we randomly sampled screening participants and intentionally selected breasts that are denser or fattier which resulted in a distribution that is not representative of the real population.” \cite{sorkhei_csaw-m_2021}. However, this type of reflexivity was largely uncommon. This was seen for datasheet 11 which labelled their data filtering techniques as “neutral” to indicate that their goal was to avoid introducing additional biases into the model. This is in contrast to the theme of ‘questioning claims of neutrality’ which embraces the situatedness of decisions. 

Overall, qualitative exploration of responses to structured dataset documentation suggests, as we observed quantitatively, a tendency on the part of dataset creators to answer even those questions that touch on reflexivity themes at minimal length and depth. At times, the answers to these questions actively use the language of neutrality and objectivity that is critiqued by proponents of reflexive work. That is: we find no compelling argument that datasheets are, at present, being used as tools for reflexivity.

\section{Discussion} \label{discussion}

Based on the lack of reflexivity within structured documentation frameworks and responses, we identify potential barriers to practicing reflexive data work, suggest approaches for increased adoption of reflexivity in the FAccT community, and provide recommendations for how structured data documentation frameworks can better support dataset developers. Lastly, we discuss potential limitations of our work. 

\subsection{Barriers to Practicing Reflexivity} \label{discussion_barriers}

Our findings point toward a meaningful disconnect: structured data documentation is touted as a way to support reflexivity in data work \cite{cambo_model_2022, gebru_datasheets_2021, miceli_documenting_2021}, but we find minimal evidence of the key elements of reflexivity in the documentation frameworks themselves or in their application. 

A major barrier to practicing reflexivity through structured  documentation may be the nature of reflexivity itself. The ‘strong objectivity’ necessary for reflexivity often comes with discomfort on the part of the dataset creators \cite{harding_after_1992,hesse-bibber_feminist_2012}. It requires researchers to articulate and engage with assumptions and norms that, within the context of a particular domain, often go tacit and unspoken \cite{klein_beyond_2021}. In data science contexts, where these tacit practices are by convention made ‘invisible’ to final data products, their intentional resurfacing in the form of reflexivity can be counter to training, institutional incentives, and long-established precedent \cite{muller_forgetting_2022, thomer_craft_2022}. Reflexivity is a skill and a challenging process, and without structural support and training, we cannot fairly expect dataset creators to engage in that process.

We can also identify the difficulty of producing good dataset documentation in general as a potential barrier to the use of structured documentation as a vehicle for reflexivity. Where, as we observe in our findings, datasheets are filled in with one-word or one-sentence answers, it is unsurprising that reflexivity cannot take place. For example, datasheet 12, despite documenting that the dataset was about people, responded “N/A” to Q17-19 (i.e., whether subpopulations or individuals were identifiable or if the data was sensitive). This would situate the lack of reflexivity in structured documentation frameworks in a growing field of scholarship concerned with the quality of data documentation itself \cite{bhardwaj_state_2024,Bhardwaj_FAccT_2024,yang_navigating_2023}. Barriers to reflexivity in this case might include a lack of adaptability and accessibility of existing documentation frameworks \cite{mcmillan-major_data_2024}, insufficient resources for the administrative burden of developing and maintaining in-depth and reflexive documentation \cite{heinke_dataset_2025}, and a desire on the part of dataset creators for automated documentation (thus sidestepping the goal of reflexivity entirely) \cite{heger_understanding_2022}.

The barriers mentioned above echo those that have been identified by previous work. Thus, at this point, it is important to consider why these issues have not yet been addressed. While structured frameworks do aim to facilitate reflexivity, there remains a lack of progress with the use of dataset documentation for this purpose. Social ethics research has described this as an “ought-is” problem, i.e., the difficulty in the implementation of an aspirational norm into regular practice \cite{Sisk_2020}. Superficial, short-lived attempts at introducing practices like reflexivity can additionally cause feelings of disempowerment. To address these barriers and their persistence, implementation science offers methods to instantiate interventions that can be implemented and disseminated broadly through collaborations between ethics experts, domain experts, and key stakeholders \cite{Sisk_2020}. This presents an opportunity for future multidisciplinary research that explores how insights from implementation science \cite{Sisk_2020} might improve the practice of reflexive documentation within the data sciences.

\subsection{Expanding Reflexivity at FAccT} \label{discussion_expanding}

Ongoing and promising efforts, among them the structured documentation frameworks we explore here and the generous exchange of strategies and ideas within the FAccT community, have made meaningful change to how reflexivity is engaged with at FAccT. We recommend next steps for that change, to address shortcomings in how reflexivity is conceptualized in the FAccT literature (RQ1). First, and most easily remedied, is the heavy reliance on Bourdieu reflexivity in FAccT publications, and then often only in the form of citation without deeper engagement. Scholars of reflexivity have made the case that Bourdieu’s conception of reflexivity, while influential and impactful, stands relatively apart from the wider discourse on reflexivity \cite{guttormsen_thinking_2023, kenway_bourdieus_2004, ribeiro_bourdieu_2022}. An expansion of the kinds of reflexivity literature informing FAccT scholarship, and of the engagement with this literature beyond citation \cite{bruton_citation_2025} could expose FAccT authors and readership to novel strategies for incorporating reflexive practice into their dataset development. 

Expanded reflexivity discussions could also support the incorporation of elements of reflexive practice not currently captured at FAccT, such as the privilege hazard. As articulated by D’Ignazio and Klein \cite{dignazio_introducing_2023}, the privilege hazard captures the idea that the scientists and practitioners working with data are often, by virtue of education and social position, poorly equipped to identify and account for inequities in their work. Engagement with the elements of reflexivity captured in our codebook requires awareness of them as topics \textit{worth} engaging with. The literature on the privilege hazard, along with work on intellectual humility and embracing the limitations of our own perspectives \cite{hoekstra_aspiring_2021,kerr_digital_2019} will support dataset creators in developing sensitivity to these topics.  

In addition to changes to individual practice, there is room for change in organizational engagement with reflexivity. Currently, the FAccT guidelines discuss the inclusion of reflexivity through an optional positionality statement. The guidelines are that this statement must be included within the endmatter section but not for the anonymous submission, i.e., positionality need only be included once the paper is accepted. Here, the norms of the FAccT community implicitly marginalize the inclusion of reflexive practice within research submissions. Firstly, the placement of positionality as an \textit{endmatter} section rather than within the main paper runs counter to the idea that reflexivity should be considered throughout the research process. Secondly, the enforcement of its inclusion only at the time of acceptance means that authors’ reflexivity on topics such as scholarly context is never considered by reviewers alongside the research methods presented. While we do not recommend placing positionality statements at the forefront or requiring them to be mandatory, there is a tension here which needs to be explored further. As FAccT’s shared blog post states, the positionality statement “does not have to include an identity disclosure” \cite{liang_reflexivity_2021}. This absence means positionality statements could be included at the time of review without compromising anonymity or diminishing the importance of reflexivity. We provide individual and collective positionality statements and reflection in the endmatter section, per FAccT guidelines.

\subsection{Recommendations for Future Practice} \label{discussion_recommendations}

In order for structured dataset documentation to more effectively enable reflexivity (RQ2 and RQ3), we make several recommendations for future practice. Of the 50 standalone questions in the datasheets template (omitting 7 questions that ask “any other comments”), we find 22 questions (44\%) to be relevant to one or more of the themes of reflexivity we identify in Section \ref{findings_themes}. Accordingly, we make recommendations about how each of the 22 datasheet questions can better prompt dataset creators to be reflexive by proposing extensions to each question. These recommendations are informed by broader discourse on reflexivity and the improvement of data practices within ML \cite{Bender_2021,Bhardwaj_FAccT_2024, boyd_datasheets_2021, miceli_documenting_2021, muller_forgetting_2022, Passi_Barocas_2019, Scheuerman_Hanna_Denton_2021,Sambasivan_2021,Thylstrup_2022}. We present the question extensions for a sample of the 22 datasheet questions in Table \ref{tbl:tbl2}. In Appendix \ref{appendixe}, we provide the full version of this table which includes further explanation of the extensions we propose.

\definecolor{Black}{rgb}{0,0,0}
\begin{longtblr}
[
  theme = mytheme,
  label = {tbl:tbl2},
  caption = {Recommendations on incorporating reflexivity in datasheet question prompts.},
]{
  width = \linewidth,
  colspec = {Q[250]Q[360]Q[313]},
  cells = {t},
  hlines,
  vlines = {Black},
  vline{1} = {-}{black},
}
Datasheet Questions (abbreviated from \cite{gebru_datasheets_2021}) & Relevant themes (identified in stage 2) & How to incorporate reflexivity into the question~\\
1. What is the purpose of the dataset? & {\labelitemi\hspace{\dimexpr\labelsep+0.5\tabcolsep}Attention to scholarly context\\\labelitemi\hspace{\dimexpr\labelsep+0.5\tabcolsep}Questioning claims of neutrality\\\labelitemi\hspace{\dimexpr\labelsep+0.5\tabcolsep}Accountability beyond transparency~} & \textit{How was this gap or task identified? Was the gap identified by the people affected by this dataset?}\\
2. Who created the dataset? & {\labelitemi\hspace{\dimexpr\labelsep+0.5\tabcolsep}Attention to scholarly context\\\labelitemi\hspace{\dimexpr\labelsep+0.5\tabcolsep}Questioning claims of neutrality\\\labelitemi\hspace{\dimexpr\labelsep+0.5\tabcolsep}Moving beyond individual positionality to consider relationality\\\labelitemi\hspace{\dimexpr\labelsep+0.5\tabcolsep}Accountability beyond transparency} & \textit{What roles and hierarchies are present on the research team? Does the entity or organization impose restrictions or limitations on dataset subjects or processes?}\\
3. Who funded the creation of the dataset? & {\labelitemi\hspace{\dimexpr\labelsep+0.5\tabcolsep}Marginalization and power~\\\labelitemi\hspace{\dimexpr\labelsep+0.5\tabcolsep}Attention to scholarly context} & \textit{Are there any limiting factors or conflicts of interest associated with the funding of their dataset?}\\
7. Does the dataset contain all possible instances or is it a sample?~ & {\labelitemi\hspace{\dimexpr\labelsep+0.5\tabcolsep}Attention to scholarly context\\\labelitemi\hspace{\dimexpr\labelsep+0.5\tabcolsep}Questioning claims of neutrality~\\\labelitemi\hspace{\dimexpr\labelsep+0.5\tabcolsep}Accountability beyond transparency} & \textit{What other sampling approaches were considered? Who is relevant but not represented? Why is this tradeoff permissible?~}\\
26. Were any ethical review processes conducted? & {\labelitemi\hspace{\dimexpr\labelsep+0.5\tabcolsep}Marginalization and power\\\labelitemi\hspace{\dimexpr\labelsep+0.5\tabcolsep}Attention to historical context\\\labelitemi\hspace{\dimexpr\labelsep+0.5\tabcolsep}Accountability beyond transparency} & \textit{What ethical concerns are important but may not be captured by these institutional processes? What standards or reviews would be applied if those offended but not involved had a say?} 
\end{longtblr}

To further engage with each theme of reflexivity, we recommend the following changes to practice and suggest resources to support these changes. 
\begin{itemize}
    \item \textbf{Marginalization and power:} Reflexive data work can engage critically with this theme by identifying power asymmetries, structurally exploitative relations implicit within data work \cite{Young_2008}, whose viewpoints are represented, and which values are embedded within and prioritized throughout the dataset development process. Berkeley’s Othering and Belonging Institute provides guidance on developing ‘power maps’ to support teams in identifying the dynamics and power relations between stakeholders involved in proposed projects \cite{othering__belonging_institute_power_2025}. Dataset developers should also consider how the data bodies lens \cite{Our_Data_Bodies_ODB} can be used during dataset planning to position data as akin to human bodies, thereby framing their own dataset development process as something with the possibility to cause tangible harms and violence \cite{Our_Data_Bodies_ODB,Young_2008}. For example, civil society organizations advocate for the treatment of civil registration data to be equal to its citizens, and thus not traded with third party organizations \cite{DeSouza_Taylor_2025}. 
    \item \textbf{Attention to historical context:} Reflexive data documentation should explicitly acknowledge social, political, or cultural conditions relevant to the data work being completed and data objects being produced. Denton et al. suggest denaturalizing datasets (which they characterize as “critical information infrastructure underpinning ML”) to make the process of their development visible within specific “socially, historically, geographically, and institutionally situated contexts”, by examining the genealogy of the dataset \cite{denton_genealogy_2021}.
    \item \textbf{Attention to scholarly context:} Reflexive data work should problematize the field (e.g., data or computer sciences) and the scholastic perspective itself (e.g., how we might construct a dataset differently for research purposes than for community use). In dataset documentation, this might take the form of critical engagement with chosen data methods and metrics, and discussion of the processes of ‘datafication’ and quantification. The ‘Toolkits for transdisciplinary research’ project \cite{bammer_toolkits_2017} compiles resources that help scholars (and data practitioners) articulate their own domain assumptions and bring them in conversation with other perspectives.
    \item \textbf{Questioning claims of neutrality:} Dataset documentation can demonstrate reflexivity and consider the situatedness of data practices by articulating researcher positionality through self-identity, research team experience, training, institutional influence, and epistemological norms and how this positionality impacts the dataset development process. Secules et al. identify six aspects of research important to reflect on within positionality statements \cite{Secules_2021} and Jacobson et al. provide a tool for developing a `positionality map' \cite{Jacobson_Mustafa_2019}. Dataset developers can also refer to examples of positionality statements and relationality within different disciplines \cite{Curran_2021} and within the dataset development in ML context \cite{bhardwaj_state_2024}. 
    \item \textbf{Moving beyond individual positionality to consider relationality:} Reflexive data work should partake in active reflexivity practices and discuss collective reflexivity. To implement more relational reflexivity, dataset developers might adapt D’Ignazio and Klein’s invisible labour iceberg as a group exercise to identify the unacknowledged or uncredited work enabling the dataset’s production \cite{dignazio_7_2020} or write and compare individual journals during dataset development to facilitate collaborative reflection on process \cite{linabary_wine_2021}.
    \item \textbf{Accountability beyond transparency:} Structured data documentation is often discussed as a tool for transparency, but this alone is not equivalent to reflexivity. The structured documentation should also demonstrate accountability beyond documenting process by considering negative downstream and long-term impacts. Dataset creators can reflect on biases that can be introduced at different stages of the dataset development process, as outlined by \cite{suresh_framework_2021}, and how to mitigate them.
\end{itemize}

\subsection{Limitations} \label{discussion_limitations}

Our findings and discussion should be considered in light of the limitations of our methodology. Our review of the literature on reflexivity adopted critical review rather than systematic review methodologies \cite{allard_little_2024,Campbell_2023,dixon-woods_conducting_2006}. These critical review strategies are ideal for synthesizing ‘big picture’ ideas across topics like reflexivity that are contested and “not consistently defined or operationalised across the field” \citep[p.~2]{dixon-woods_conducting_2006}, but do mean that our reflexivity codebook is based on a selection of the literature rather than on a hypothetical, comprehensive body of all work on reflexivity.

We review only a sample of structured dataset documentation templates and completed documents. These samples are based on field standards (in the case of the chosen templates \cite{mcmillan-major_data_2024}) and prior peer-reviewed scholarship (in the case of the completed documents \cite{bhardwaj_state_2024}), but are not comprehensive. It is possible that other bodies of dataset documentation engage differently with reflexivity. Our analysis of completed dataset documentation from NeurIPS also focused on the structured documents themselves, usually included as an appendix. It is possible that further reflexivity was discussed in the content of the main papers, or elsewhere in the supplementary material. 

Our study advocates for more reflexivity in documentation (e.g., to surface power imbalances) and provides recommendations to improve reflexivity in documentation practices. However, it does not provide mechanisms for accountability or for doing reflexivity at the institutional level (e.g., address root causes of core power imbalances). This would be a critical next step for future research and would make the process of documentation more empowering, giving data developers and users the means to act on the results of such documentation.

\section{Conclusion} \label{conclusion}

Through mixed-method thematic analysis and corpus-assisted discourse analysis, we explored the reality of structured dataset documentation frameworks as tools for reflexivity. By identifying key themes in how reflexivity is conceived in the FAccT literature and beyond, and empirically demonstrating the general lack of engagement with these themes in both structured documentation frameworks and published applications thereof, we make the case that at present, structured documentation is not contributing to reflexive data work in the ways intended by creators and users. This work is intended to serve as a ‘critical friend’ \cite{becker_insolvent_2023} to existing scholarship on reflexive data work and dataset documentation frameworks by identifying evidence-based, specific issues with current practice, and making actionable recommendations for improvement. We emphasize the necessity of positioning reflexivity in dataset documentation as a domain-wide, generative process: just as project-level reflexivity must be relational, domain-level reflexivity requires the commitment of researchers and practitioners across the data sciences.


\section*{Acknowledgments}

This research is partially supported by NSERC through RGPIN-2016-06640 and RGPIN-2025-07063, the Canada Foundation for Innovation, and the Ontario Research Fund.

\section*{Generative AI Disclosure Statement}

This paper was developed and written without any assistance from generative AI tools. 

\section*{Author Contributions}

This research was developed and written equally by both first authors. The last author contributed in a supervisory rule, such as by reviewing, editing, and advising the research. 

\section*{Reflexivity}
Below we discuss our reflexive process for this research project, structured according to the six themes we presented in our study. We hope that by demonstrating how we approached reflexivity, other researchers at FAccT and otherwise can benefit from our example as a useful resource.

\subsubsection*{Questioning Claims of Neutrality (Individual Positionality Statements)} \label{positionality}\hfill\break
\textbf{Eshta (she/her):} I am a PhD candidate at University of Toronto’s Faculty of Information. My prior academic background is rooted in information technology and data science, but my doctoral work is within a sociotechnical space. My perspective has thus shifted from seeing technology as a neutral entity, a tool that must be used and advanced further, to recognition of the need for a critical approach to technology that takes a reflexive, ethics, and justice based lens. My research partly examines the intrinsic bias that dataset developers impart on their datasets and my view on the importance of reflexivity is informed by this. 

\noindent\textbf{Ciara:} Ciara Zogheib (she/her) is a PhD candidate at the University of Toronto’s Faculty of Information, with a research focus on the information practices of data work. She has professional experience as a data scientist and a commitment to mixed-methods and critical scholarship, both of which inform the way she researches data and data practice.

\noindent\textbf{Christoph:} Christoph Becker (he/him) is a Professor of Information at the University of Toronto, has long done research in data curation, and sometimes describes himself as a 'recovered computer scientist' because he retroactively realized that his computer science education instilled technosolutionist beliefs that took years to unlearn. His perspective on reflexivity is influenced by intersectional feminists and critical systems thinkers who both emphasize the inevitable situatedness of knowledge claims, the politics of such claims, and the inevitable gaps in them. 

\subsubsection*{Moving beyond Individual Positionality to Consider Relationality (Group Positionality)} \label{grouppositionality}

As a group, two of the authors (CZ and EB) are PhD candidates at the same stage in our studies, while CB is in a senior role as a faculty member and EB’s doctoral supervisor. We have previously collaborated on research projects, giving us a level of familiarity with each other’s preferred ways of working and differing expertise. All three authors are affiliated with the University of Toronto’s Faculty of Information. The freedom to explore our chosen topic, the availability of funding to carry out this project, and the credibility associated with our institution are privileges that influenced the completion of this work. During the project, we incorporated active reflexivity practices into our collaboration, including: frequently discussing our individual perspectives and positions, establishing group norms of productively questioning assumptions or misunderstandings, and explicitly negotiating allocation of work and credit.

\subsubsection*{Attention to Historical Context} \label{historical}

In the present academic climate, research that critiques or questions equity in dataset development processes, reflects on the treatment of marginalized communities during data collection, or even uses the term ‘feminist’ has the possibility of not being funded, and worse yet, can impact the ability of researchers to work safely. As scholars at a Canadian institution, we recognize our privilege to conduct this type of research freely.

\subsubsection*{Attention to Scholarly Context} \label{scholarly}

Our scholarly context has shaped our ability to do this work: at present, reflexivity and positionality are considered valid research topics in data and CS communities, which was not always the case; these communities have also relatively recently become more open to incorporating perspectives from queer and feminist theory, as we do in our review of the literature. Scholarly context has also shaped how we do this work: by engaging with reflexivity as a subject of academic discourse (e.g., reviewing academic books and articles), we potentially overlook non-academic conceptions of reflexivity, positionality, identity etc. Our research question binds us to a particular body of discourse and particular conceptions of reflexivity. We reflect by asking: Who has been excluded from these traditional publishing venues? How might reflexivity as an academic concept differ from reflexivity as practiced ‘on the ground’?

\subsubsection*{Marginalization and Power} \label{marginalization}

We use power mapping techniques \cite{othering__belonging_institute_power_2025} to surface power asymmetries within our project. Instead of developing a visual representation of our power map, we use the power analysis prompts \cite{othering__belonging_institute_power_2025}. The first step of developing a power map is articulating the goal of our analysis. We define our goal as: examining and improving reflexivity in data documentation. Below we identify relevant actors and how they may be impacted directly/indirectly by our goal, support our goal, and/or oppose our goal.  
\begin{itemize}
    \item Dataset developers: These actors have a moderate to high interest in reflexivity in data documentation as they are the ones responsible for the documentation and are the core audience for this research. Dataset developers who are not researchers or members of the FAccT community would not have as much agency and power to provide direct feedback to this work. Finally, they may provide less support to this goal since recommendations to improve reflexivity require increased labour by them. 
    
    \item Researchers interested in dataset documentation: These actors have high interest because of the relevance of the topic to their own research. They additionally have the ability to impact or influence the topic as we submit and present our research at a conference and thus actively seek their feedback and evaluation. Researchers would likely provide support for this work as they are already involved in similar dataset documentation research. 

    \item Dataset users: These actors have moderate to high interest since improved documentation standards makes dataset use easier and more informed. If these actors are not also researchers or members of the FAccT community, they would have less agency or power to provide direct feedback to this goal. Dataset users would likely provide a high amount of support as greater reflexivity would enable more effective use of datasets. 

    \item People or communities described in datasets: These actors may have low interest in this goal because they are not actively engaged in the dataset development process. However, they would have high interest because they would benefit from any uptake of the recommendations in improving reflexivity in data work. The actors however have low power since we did not seek out people or communities described in datasets to obtain their feedback or perspectives. These actors could provide high support as more reflexive dataset development could translate to more equitable data collections processes. 

\end{itemize}

\subsubsection*{Accountability beyond Transparency} \label{accountability}

We support transparency by including appendices showing our work, and this positionality section articulating our intellectual commitments and biases. We consider potential negative, downstream impacts of this work as part of further accountability, including: (1) By asking people to consider reflexivity in the depth we recommend, we ask them to do additional labour for which not everyone has the funding or time; (2) Not everyone has a professional environment that supports personal disclosure in the name of reflexivity; (3) Our conceptions of reflexivity are based on review of English-language publications, most commonly by authors affiliated with Western institutions; by using these conceptions to review existing dataset publications, we’re arguably imposing western standards on scholars around the world working in different cultural contexts; and (4) We assume a level of agency over dataset development that not all developers have (e.g., if they are working in an organizational context that has specific guidelines they have to follow). 

We address or mitigate these potential negatives as follows: (1) The reflexivity resources we recommend are open online (no access fees); it is additional labour (which should be acknowledged and compensated) but we provide a variety of recommendations and steps that can be incorporated throughout the dataset development process and emphasize that given the current state of reflexivity in this domain, any efforts are commendable; (2) Most of our recommendations move away from the ‘mere’ listing of individual identities by providing ways for developers to incorporate reflexivity even if disclosure isn’t safe or comfortable; (3) We acknowledge that we are limited by our language fluency and access to particular resources; a similar review of other conceptions of reflexivity and positionality from other cultural perspectives or literature in other languages would be appreciated and a valuable contribution; and (4) Instead of making one mandatory set of steps, we provide a range of options and resources so that people can pursue reflexivity using the approach most accessible to them.

\bibliographystyle{ACM-Reference-Format}
\bibliography{facct26bib_part1,facct26bib_appendix,facct26bib_eshta,facct26bib_pascalian,facct26bib_RTA}

\newpage

%
\begin{appendix}  \label{appendix}
\section*{Appendix}

\section{Reflexivity Literature Corpus} \label{appendixa}

We provide a list of all FAccT literature (n = 84) that mention reflexivity (reflexiv-) across all archived years below in Table \ref{tab:tab3}. Notably there is an increase of mentions of reflexivity from 2021 (4 papers) to 2022 (18 papers). From 2022 onwards, this number remains relatively constant. 

Our reflexivity literature corpus consisted of 30 documents (books, journal articles, and conference proceedings) from feminist and poststructuralist reflexivity \cite{bardzell_towards_2011, dignazio_introducing_2023, england_getting_1994, haraway_situated_1988,harding_after_1992,harding_rethinking_1992,hesse-bibber_feminist_2012,ravera_reflexivity_2023,soedirgo_toward_2020,suchman_located_2002}, Bourdesian reflexivity \cite{bourdieu_pascalian_2000,bourdieu_science_2004,hess_neoliberalism_2013,kenway_bourdieus_2004,maton_reflexivity_2003}, critical data studies \cite{boyd_quinta_2023,cambo_model_2022,kraft_knowledge-enhanced_2024,miceli_documenting_2021,schroeder_disclosure_2025,tanweer_why_2021}, broader methods literature \cite{day_reflexive_2012,finlay_negotiating_2002, jamieson_reflexivity_2023,folkes_moving_2023,mccorkel_what_2003,pillow_confession_2003}, and others \cite{emirbayer_race_2012, hirsbrunner_critical_2024, seaver_vertesi_knowing_2019}.

\begin{small}
\begin{longtable}{|>{\hspace{0pt}}m{0.630\linewidth}|>{\hspace{0pt}}m{0.300\linewidth}|} 
\caption{FAccT literature that mention reflexivity across all archived years.\label{tab:tab3}}\\ 
\hline
Paper & DOI \endfirsthead 
\hline
Reflexive Prompt Engineering: A Framework for Responsible  Prompt Engineering and AI Interaction Design. & doi.org/10.1145/3715275.3732118 \\ 
\hline
Four Years of FAccT: A Reflexive, Mixed-Methods Analysis of Research  Contributions, Shortcomings, and Future Prospects. & doi.org/10.1145/3531146.3533107 \\ 
\hline
Documenting Computer Vision Datasets: An Invitation to  Reflexive Data Practices. & doi.org/10.1145/3442188.3445880 \\ 
\hline
Disclosure without  Engagement: An Empirical Review of Positionality Statements at FAccT. & doi.org/10.1145/3715275.3732079 \\ 
\hline
Who Gets Heard? Calling Out  the "Hard-to-Reach" Myth for Non-WEIRD Populations' Recruitment and  Involvement in Research. & doi.org/10.1145/3715275.3732055 \\ 
\hline
Smallset  Timelines: A Visual Representation of Data Preprocessing Decisions. & doi.org/10.1145/3531146.3533175 \\ 
\hline
Constructing Capabilities: The Politics of Testing  Infrastructures for Generative AI. & doi.org/10.1145/3630106.3659009 \\ 
\hline
Skin Deep: Investigating  Subjectivity in Skin Tone Annotations for Computer Vision Benchmark Datasets. & doi.org/10.1145/3593013.3594114 \\ 
\hline
Towards  Labor Transparency in Situated Computational Systems Impact Research. & doi.org/10.1145/3593013.3594060 \\ 
\hline
Augmented  Datasheets for Speech Datasets and Ethical Decision-Making. & doi.org/10.1145/3593013.3594049 \\ 
\hline
Fairness Beyond the Algorithmic Frame: Actionable Recommendations for an  Intersectional Approach. & doi.org/10.1145/3715275.3732210 \\ 
\hline
Agent Allocation  of Moral Decisions in Human-Agent Teams: Raise Human Involvement and Explain  Potential Consequences. & doi.org/10.1145/3715275.3732157 \\ 
\hline
Frameworks, Methods and Shared Tasks:  Connecting Participatory AI to Trustworthy AI Through a Systematic Review of  Global Projects. & doi.org/10.1145/3715275.3732148 \\ 
\hline
Misabstraction in Sociotechnical Systems. & doi.org/10.1145/3715275.3732122 \\ 
\hline
Hidden Conflicts in Neural Networks and their Implications for  Explainability. & doi.org/10.1145/3715275.3732100 \\ 
\hline
A Trustworthiness-based Metaphysics of Artificial  Intelligence Systems. & doi.org/10.1145/3715275.3732091 \\ 
\hline
Opening the Scope of Openness in  AI. & doi.org/10.1145/3715275.3732087 \\ 
\hline
Labor, Power, and Belonging: The Work of Voice in the Age of AI  Reproduction.~ & doi.org/10.1145/3715275.3732082 \\ 
\hline
From Lived Experience to Insight: Unpacking the  Psychological Risks of Using AI Conversational Agents. & doi.org/10.1145/3715275.3732063 \\ 
\hline
Rethinking AI Safety:  Provocations from the History of Community-based Practices of Road and Driver  Safety. & doi.org/10.1145/3715275.3732062 \\ 
\hline
Liberatory  Collections and Ethical AI: Reimagining AI Development from Black Community  Archives and Datasets. & doi.org/10.1145/3715275.3732058 \\ 
\hline
~Pragmatic  Fairness: Evaluating ML Fairness Within the Constraints of Industry. & doi.org/10.1145/3715275.3732040 \\ 
\hline
The Gaps that Never Were: Reconsidering  Responsible AI's Principle-Practice Problem. & doi.org/10.1145/3715275.3732024 \\ 
\hline
African Data Ethics: A Discursive Framework for Black  Decolonial AI. & doi.org/10.1145/3715275.3732023 \\ 
\hline
Examining the Expanding Role of Synthetic Data Throughout the AI Development  Pipeline. & doi.org/10.1145/3715275.3732005 \\ 
\hline
On the Quest for Effectiveness in  Human Oversight: Interdisciplinary Perspectives. & doi.org/10.1145/3630106.3659051 \\ 
\hline
Data,  Annotation, and Meaning-Making: The Politics of Categorization in Annotating  a Dataset of Faith-based Communal Violence. & doi.org/10.1145/3630106.3659030 \\ 
\hline
Balancing Act: Evaluating People's Perceptions of Fair  Ranking Metrics. & doi.org/10.1145/3630106.3659018 \\ 
\hline
Actionable Recourse for Automated Decisions: Examining the Effects of  Counterfactual Explanation Type and Presentation on Lay User Understanding. & doi.org/10.1145/3630106.3658997 \\ 
\hline
Animation and Artificial Intelligence. & doi.org/10.1145/3630106.3658995 \\ 
\hline
Participation in the age of foundation models. & doi.org/10.1145/3630106.3658992 \\ 
\hline
Learning about Responsible AI On-The-Job: Learning  Pathways, Orientations, and Aspirations. & doi.org/10.1145/3630106.3658988 \\ 
\hline
Knowledge-Enhanced Language Models Are Not  Bias-Proof: Situated Knowledge and Epistemic Injustice in AI. & doi.org/10.1145/3630106.3658981 \\ 
\hline
Epistemic Power in AI Ethics Labor: Legitimizing Located  Complaints. & doi.org/10.1145/3630106.3658973 \\ 
\hline
Machine learning data practices through a data  curation lens: An evaluation framework. & doi.org/10.1145/3630106.3658955 \\ 
\hline
From Model Performance to Claim: How a Change of Focus in Machine  Learning Replicability Can Help Bridge the Responsibility Gap. & doi.org/10.1145/3630106.3658951 \\ 
\hline
The Harmful Fetishisation of Reductive Personal  Tracking Metrics in Digital Systems. & doi.org/10.1145/3630106.3658943 \\ 
\hline
AI Failure Cards:  Understanding and Supporting Grassroots Efforts to Mitigate AI Failures in  Homeless Services. & doi.org/10.1145/3630106.3658935 \\ 
\hline
The Digital Faces of Oppression and Domination: A Relational  and Egalitarian Perspective on the Data-driven Society and its Regulation. & doi.org/10.1145/3630106.3658934 \\ 
\hline
Insights From Insurance for Fair  Machine Learning.~ & doi.org/10.1145/3630106.3658914 \\ 
\hline
~"Like rearranging deck chairs on the Titanic"? Feasibility,  Fairness, and Ethical Concerns of a Citizen Carbon Budget for Reducing CO2  Emissions. & doi.org/10.1145/3630106.3658904 \\ 
\hline
Queer In AI: A Case Study in Community-Led  Participatory AI. & doi.org/10.1145/3593013.3594134 \\ 
\hline
Bias as Boundary Object:  Unpacking The Politics Of An Austerity Algorithm Using Bias Frameworks. & doi.org/10.1145/3593013.3594120 \\ 
\hline
Representation,  Self-Determination, and Refusal: Queer People's Experiences with Targeted  Advertising. & doi.org/10.1145/3593013.3594110 \\ 
\hline
What's fair is... fair? Presenting JustEFAB, an ethical  framework for operationalizing medical ethics and social justice in the  integration of clinical machine learning: JustEFAB. & doi.org/10.1145/3593013.3594096 \\ 
\hline
(Anti)-Intentional Harms: The Conceptual  Pitfalls of Emotion AI in Education. & doi.org/10.1145/3593013.3594088 \\ 
\hline
An Empirical  Analysis of Racial Categories in the Algorithmic Fairness Literature. & doi.org/10.1145/3593013.3594083 \\ 
\hline
A Systematic Review of Ethics Disclosures in  Predictive Mental Health Research/ & doi.org/10.1145/3593013.3594082 \\ 
\hline
I'm fully who I am: Towards  Centering Transgender and Non-Binary Voices to Measure Biases in Open  Language Generation. & doi.org/10.1145/3593013.3594078 \\ 
\hline
We try to empower them - Exploring Future Technologies to  Support Migrant Jobseekers. & doi.org/10.1145/3593013.3594056 \\ 
\hline
Algorithmic Decisions, Desire  for Control, and the Preference for Human Review over Algorithmic Review. & doi.org/10.1145/3593013.3594041 \\ 
\hline
You Sound Depressed: A  Case Study on Sonde Health's Diagnostic Use of Voice Analysis AI. & doi.org/10.1145/3593013.3594032 \\ 
\hline
Striving for Affirmative Algorithmic Futures: How the Social  Sciences can Promote more Equitable and Just Algorithmic System Design. & doi.org/10.1145/3593013.3594022 \\ 
\hline
AI's Regimes  of Representation: A Community-centered Study of Text-to-Image Models in  South Asia. & doi.org/10.1145/3593013.3594016 \\ 
\hline
Rethinking Transparency as a Communicative  Constellation. & doi.org/10.1145/3593013.3594010 \\ 
\hline
Envisioning  Equitable Speech Technologies for Black Older Adults. & doi.org/10.1145/3593013.3594005 \\ 
\hline
Making Intelligence: Ethical Values  in IQ and ML Benchmarks.~ & doi.org/10.1145/3593013.3593996 \\ 
\hline
WEIRD FAccTs: How Western, Educated, Industrialized, Rich, and  Democratic is FAccT? & doi.org/10.1145/3593013.3593985 \\ 
\hline
Fairness in machine learning from the perspective of  sociology of statistics: How machine learning is becoming scientific by  turning its back on metrological realism. & doi.org/10.1145/3593013.3593974 \\ 
\hline
From Demo to Design in Teaching Machine  Learning. & doi.org/10.1145/3531146.3534634 \\ 
\hline
Algorithmic Tools in Public  Employment Services: Towards a Jobseeker-Centric Perspective. & doi.org/10.1145/3531146.3534631 \\ 
\hline
Theories of \`{}Gender'  in NLP Bias Research.~ & doi.org/10.1145/3531146.3534627 \\ 
\hline
Limits and  Possibilities for \`{}Ethical AI' in Open Source: A Study of Deepfakes. & doi.org/10.1145/3531146.3533779 \\ 
\hline
CounterFAccTual: How FAccT Undermines Its  Organizing Principles. & doi.org/10.1145/3531146.3533241 \\ 
\hline
~How  Platform-User Power Relations Shape Algorithmic Accountability: A Case Study  of Instant Loan Platforms and Financially Stressed Users in India. & doi.org/10.1145/3531146.3533237 \\ 
\hline
Evaluation Gaps in Machine Learning Practice. & doi.org/10.1145/3531146.3533233 \\ 
\hline
Critical Tools for  Machine Learning: Working with Intersectional Critical Concepts in Machine  Learning Systems Design. & doi.org/10.1145/3531146.3533207 \\ 
\hline
Accountable Data: The Politics and Pragmatics of Disclosure  Datasets. & doi.org/10.1145/3531146.3533201 \\ 
\hline
The Algorithmic  Imprint. & doi.org/10.1145/3531146.3533186 \\ 
\hline
Evidence for Hypodescent  in Visual Semantic AI. & doi.org/10.1145/3531146.3533185 \\ 
\hline
German AI Start-Ups and 'AI Ethics':  Using A Social Practice Lens for Assessing and Implementing Socio-Technical  Innovation. & doi.org/10.1145/3531146.3533156 \\ 
\hline
~Towards Intersectional Feminist and Participatory ML: A Case Study in  Supporting Feminicide Counterdata Collection. & doi.org/10.1145/3531146.3533132 \\ 
\hline
Don't let Ricci v. DeStefano Hold  You Back: A Bias-Aware Legal Solution to the Hiring Paradox.~ & doi.org/10.1145/3531146.3533129 \\ 
\hline
Mind the Gap: Autonomous Systems, the Responsibility Gap,  and Moral Entanglement. & doi.org/10.1145/3531146.3533106 \\ 
\hline
The Values Encoded in Machine Learning Research. & doi.org/10.1145/3531146.3533083 \\ 
\hline
The Ethics of Emotion in Artificial Intelligence  Systems. & doi.org/10.1145/3442188.3445939 \\ 
\hline
PAn Action-Oriented AI Policy Toolkit for Technology Audits  by Community Advocates and Activists. & doi.org/10.1145/3442188.3445938 \\ 
\hline
You Can't  Sit With Us: Exclusionary Pedagogy in AI Ethics Education & doi.org/10.1145/3442188.3445914 \\ 
\hline
Closing the AI accountability gap: defining an end-to-end framework for  internal algorithmic auditing.~ & doi.org/10.1145/3351095.3372873 \\ 
\hline
From ethics washing to ethics bashing: a view on tech ethics  from within moral philosophy. & doi.org/10.1145/3351095.3372860 \\ 
\hline
Studying up:  reorienting the study of algorithmic fairness around issues of power. & doi.org/10.1145/3351095.3372859 \\ 
\hline
Towards a  critical race methodology in algorithmic fairness. & doi.org/10.1145/3351095.3372826 \\ 
\hline
Racial categories in machine learning. & doi.org/10.1145/3287560.3287575 \\ 
\hline
Explaining Explanations  in AI. & doi.org/10.1145/3287560.3287574 \\
\hline
\end{longtable}
\end{small}

\section{List of Responses} \label{appendixb}

\begin{small}
\definecolor{Black}{rgb}{0,0,0}
\begin{longtblr}[
  theme = mytheme,
  label = {tbl:tbl4},
  caption = {List of responses analyzed using CADA approach.},
]{
  width = \linewidth,
  colspec = {Q[102]Q[94]Q[562]Q[110]Q[73]},
  hlines,
  vlines = {Black},
  vline{1} = {-}{black},
}

Datasheet Number & Type of Template & Paper Title                                                                                                                      & Dataset Abbreviation & Reference \\
1                & Datasheet        & Programming Puzzles                                                                                                              & program\newline\_puzzles     & \cite{schuster_programming_2021}    \\
2                & Datasheet        & SciGen: a Dataset for Reasoning-Aware Text Generation from Scientific Tables                                                     & scigen               & \cite{moosavi_scigen_2021}    \\
3                & Datasheet~       & CEDe: A collection of expert-curated datasets with atom-level entity annotations for Optical Chemical Structure Recognition      & cede                 & \cite{hormazabal_cede_2022}    \\
4                & Datasheet        & LoveDA: A Remote Sensing Land-Cover Dataset for Domain Adaptive Semantic Segmentation~                                           & loveda               & \cite{wang_loveda_2021}    \\
5                & Datasheet        & Change Event Dataset for Discovery from Spatio-temporal Remote Sensing Imagery                                                   & change\newline\_event        & \cite{mall_change_2022}    \\
6                & Datasheet        & CAESAR: An Embodied Simulator for Generating Multimodal Referring Expression Datasets                                            & caesar               & \cite{islam_caesar_2022}    \\
7                & Datasheet        & BubbleML: A Multiphase Multiphysics Dataset and Benchmarks for Machine Learning                                                  & bubbleML             & \cite{hassan_bubbleml_2023}    \\
8                & Datasheet~       & DataComp: In search of the next generation of multimodal datasets                                                                & datacomp             & \cite{gadre_datacomp_2023}    \\
9                & Datasheet        & The CPD Data Set: Personnel, Use of Force, and Complaints in the Chicago Police Department~                                      & cpd                  & \cite{horel_cpd_2021}    \\
10               & Datasheet        & The Tufts fNIRS Mental Workload Dataset  Benchmark for Brain-Computer Interfaces that Generalize                                 & tufts                & \cite{huang_tufts_2021}    \\
11               & Datasheet        & The RefinedWeb Dataset for Falcon LLM: Outperforming Curated Corpora with Web Data Only~                                         & refinedweb           & \cite{penedo_refinedweb_2023}    \\
12               & Datasheet        & VisAlign: Dataset for Measuring the Alignment between AI and Humans in Visual Perception                                         & visalign             & \cite{lee_visalign_2023}    \\
13               & Datasheet~       & American Stories: A Large-Scale Structured Text Dataset of Historical U.S. Newspapers~                                           & amerstories          & \cite{dell_american_2023}    \\
14               & Datasheet~~      & KeSpeech: An Open Source Speech Dataset of Mandarin and Its Eight Subdialects                                                    & kespeech             & \cite{tang_kespeech_2021}    \\
15               & Datasheet        & FLAIR: a Country-Scale Land Cover Semantic Segmentation Dataset From Multi-Source Optical Imagery                                & flair                & \cite{garioud_flair_2023}    \\
16               & Datasheet        & MedSat: A Public Health Dataset for England Featuring Medical Prescriptions and Satellite Imagery                                & medsat               & \cite{scepanovic_medsat_2023}    \\
17               & Datasheet        & PUG: Photorealistic and Semantically Controllable Synthetic Data for Representation Learning~                                    & pug                  & \cite{bordes_pug_2023}     \\
18               & Datasheet~       & Constructing a Visual Dataset to Study the Effects of Spatial Apartheid in South Africa                                          & spatial\_apart       & \cite{sefala_constructing_2021}    \\
19               & Datasheet~       & A Spoken Language Dataset of Descriptions for Speech-Based Grounded Language Learning~                                           & spoken\_lang         & \cite{kebe_spoken_2021}    \\
20               & Datasheet~       & Ambiguous Images With Human Judgments for Robust Visual Event Classification~                                                    & ambiguous            & \cite{sanders_ambiguous_2022}    \\
21               & Datasheet        & SCAMPS: Synthetics for Camera Measurement of Physiological Signals                                                               & scamps               & \cite{mcduff_scamps_2022}    \\
22               & Datasheet        & Objaverse-XL: A Universe of 10M+ 3D Objects                                                                                      & objaverse            & \cite{deitke_objaverse-xl_2023}    \\
23               & Datasheet~       & CREAK: A Dataset for Commonsense Reasoning over Entity Knowledge                                                                 & creak                & \cite{onoe_creak_2021}    \\
24               & Data Statement   & CLEVRER-Humans: Describing Physical and Causal Events the Human Way                                                              & clevrer              & \cite{mao_clevrer-humans_2022}   \\
25               & Datasheet~       & OpenProteinSet: Training data for structural biology at scale                                                                    & openprotein          & \cite{ahdritz_openproteinset_2023}     \\
26               & Datasheet        & PROSPECT: Labeled Tandem Mass Spectrometry Dataset for Machine Learning in Proteomics                                            & prospect             & \cite{shouman_prospect_2022}    \\
27               & Datasheet~       & MADLAD-400: A Multilingual And Document-Level Large Audited Dataset                                                              & madlad               & \cite{kudugunta_madlad-400_2023}    \\
28               & Datasheet~       & STAR: A Benchmark for Situated Reasoning in Real-World Videos                                                                    & star                 & \cite{wu_star_2021}    \\
29               & Datasheet~       & BiToD: A Bilingual Multi-Domain Dataset For Task-Oriented Dialogue Modeling                                                      & bitod                & \cite{lin_bitod_2021}    \\
30               & Datasheet~       & RenderMe-360: A Large Digital Asset Library and Benchmarks Towards High-fidelity Head Avatars                                    & renderme             & \cite{pan_renderme-360_2023}    \\
31               & Datasheet        & ConfLab: A Data Collection Concept, Dataset, and Benchmark for Machine Analysis of Free-Standing Social Interactions in the Wild & conflab              & \cite{raman_conflab_2022}    \\
32               & Datasheet ~      & CSAW-M: An Ordinal Classification Dataset for Benchmarking Mammographic Masking of Cancer                                        & csaw                 & \cite{sorkhei_csaw-m_2021}    \\
33               & Datasheet~       & WikiChurches: A Fine-Grained Dataset of Architectural Styles with Real-World Challenges                                          & wikichurches         & \cite{barz_wikichurches_2021}     \\
34               & Datasheet~       & Mathematical Capabilities of ChatGPT                                                                                             & mathgpt              & \cite{frieder_mathematical_2023}    \\
35               & Datasheet~       & COVID-19 Sounds: A Large-Scale Audio Dataset for Digital Respiratory Screening                                                   & covid                & \cite{xia_covid-19_2021}    \\
36               & Datasheet        & Addressing Resource Scarcity across Sign Languages with Multilingual Pretraining and Unified-Vocabulary Datasets                 & sign\_lang           & \cite{nc_addressing_2022}    
\end{longtblr}
\end{small}

\section{Extended Codebook} \label{appendixc}

\definecolor{Black}{rgb}{0,0,0}
\begin{longtblr}[
  theme = mytheme,
  caption = {Extended codebook of themes elicited through RTA that summarize the conceptualization of reflexivity within the compiled corpus (extension of Table \ref{tbl:tbl1}).},
]{
  width = \linewidth,
  colspec = {Q[150]Q[792]},
  row{1} = {c},
  row{5} = {c},
  row{9} = {c},
  row{13} = {c},
  row{17} = {c},
  row{21} = {c},
  cell{1}{1} = {c=2}{0.942\linewidth},
  cell{5}{1} = {c=2}{0.942\linewidth},
  cell{9}{1} = {c=2}{0.942\linewidth},
  cell{13}{1} = {c=2}{0.942\linewidth},
  cell{17}{1} = {c=2}{0.942\linewidth},
  cell{21}{1} = {c=2}{0.942\linewidth},
  hlines,
  vlines = {Black},
  vline{1} = {-}{black},
}
Theme: Marginalization and power~  & \\
Description  & This theme disentangles the roles assigned within knowledge production and how their placements within social and/or scholastic hierarchies result in unequal impacts.~  \\
Relevance to structured documentation & Reflexive data work engages critically with this theme by identifying power asymmetries, whose viewpoints are represented, and which values are embedded within and prioritized throughout the dataset development process.~  \\
Examples  & {\textit{“A reflexive analysis of power thus critiques the tendency to understand our research subjects as a unitary group. Due to a tradition of “studying down,” there is a tendency in qualitative methodology to erase hierarchies among those we study, which prevents us from understanding the heterogeneity of our participants and blinds us to the existence of power differentials between and among them” }\citep[p.~6]{day_reflexive_2012}.\\\textit{“Power differentials become evident when deciding which data to collect, how to classify it, and how to label it. Many datasets are produced with a specific computer vision product in mind. Dataset design begins as the expected outcome of that product (in terms of computational output but also of revenue) is transformed into task instructions for data collectors and annotators.” }\citep[p.~165]{miceli_documenting_2021}.\\\textit{“Testimonial injustice is a consequence of identity prejudice: We usually assign credibility automatically to speakers, and in this unreflective process, identity prejudice can unjustly lead us to grant less credibility to some speakers, typically from marginalized groups.” }\citep[p.~1440]{kraft_knowledge-enhanced_2024}.}   \\
Theme: Attention to historical context   &   \\
Description & This theme positions research as situated historically, conducted in a particular time and place with accompanying social, political, and cultural context. Reflexive research does not obscure or deny this historicity.  \\
Relevance to structured documentation  & Reflexive data documentation explicitly acknowledges social, political, or cultural conditions relevant to the data work being completed and data objects being produced. \\
Examples  & {\textit{``Can truth survive radical historicization? In other words, is the necessity of logical truths compatible with recognition of their historicity?'' } \citep[p.~2]{bourdieu_science_2004}. \\\textit{“Researchers are impacted and possess their own lived experiences (i.e., cultures, privileges, social locations) and feelings, and their impacts on the research process cannot be ignored.” }\citep[p.~3]{boyd_quinta_2023}.\\\textit{“If we are to consider that expertise and experience inform one’s data vision, then we should also consider the social, cultural and political context of this expertise and experience. In many ways, the concerns that feminist scholars address with concepts of situated knowledge and positionality are mirrored in recent trends within machine learning and data science.” }\citep[p.~4]{cambo_model_2022}.}  \\
Theme: Attention to scholarly context &  \\
Description  & This theme emphasizes how academic disciplines and research domains constitute fields with norms, incentives, and internal epistemological and evidentiary standards (all of which can shape scholarly work).   \\
Relevance to structured documentation & Reflexive data work problematizes the field (e.g., data or computer sciences) and the scholastic perspective itself (e.g., how we might construct a dataset differently for research purposes than for community use). In dataset documentation, this might take the form of critical engagement with chosen data methods and metrics, and discussion of the processes of ‘datafication’ and quantification.~   \\
Examples  & {\textit{“The field of sociology thus produces its own intellectual dispositions and it is these and the epistemic history and unconscious of the field that must be interrogated, rather than the apparently idiosyncratic view points of the individual researcher. It is via this process that the reflexive researcher can uncover and systematically explore the ‘unthought categories of thought which delimit the thinkable and predetermine the thought’.” }\citep[p.~528]{kenway_bourdieus_2004}.\\\textit{“As STS researchers begin to study the problems associated with neoliberalism, science, and technology, there is a need for a reflexive inquiry into the underlying conceptual frameworks of the STS field itself and the possibility that the some of the dominant conceptual frameworks of the field are inflected by decades of neoliberal thought.” }\citep[p.~178]{hess_neoliberalism_2013}.\\\textit{“However, the methods and norms in the disciplines are too weak to permit researchers systematically to identify and eliminate from the results of research those social values, interests, and agendas that are shared by the entire scientific community or virtually all of it.” }\citep[p.~52]{harding_rethinking_1992}.} \\
Theme: Questioning claims of neutrality  &  \\
Description & This theme rejects the notion of neutrality or objectivity within research and science. Claims of neutrality are opposed with embracing situatedness and claiming it a stronger form of objectivity, i.e., reflexivity makes research more objective and thus more legitimate by acknowledging situatedness.~  \\
Relevance to structured documentation  & Reflexive data work engages critically with the situatedness of data practices by articulating researcher positionality through self-identity, research team experience and training, institutional influence, and epistemological norms and how this positionality impacts the dataset development process.  \\
Examples  & {\textit{“Strong objectivity would specify strategies to detect social assumptions that (a) enter research in the identification and conceptualization of scientific problems and the formation of hypotheses about them (the “context of discovery”), (b) tend to be shared by observers designated as legitimate ones, and thus are significantly collective, not individual, values and interests, and (c) tend to structure the institutions and conceptual schemes of disciplines. These systematic procedures would also be capable of (d) distinguishing between those values and interests that block the production of less partial and distorted accounts of nature and social relations (“less false” ones) and those- such as fairness, honesty, detachment, and, we should add, advancing democracy- that provide resources for it.” }\citep[p.~580]{harding_after_1992}.\\\textit{“The moral is simple: only partial perspective promises objective vision. All Western cultural narratives about objectivity are allegories of the ideologies governing the relations of what we call mind and body, distance and responsibility. Feminist objectivity is about limited location and situated knowledge,not about transcendence and splitting of subject and object.” }\citep[p.~583]{haraway_situated_1988}.\\\textit{“Scientific insight comes by way of rigorous reflexivity and is not the inevitable result of one’s position in social space. The notion that white scholars, strictly because of their whiteness, are blind to certain dimensions of racial domination, while nonwhite scholars, strictly because of their nonwhiteness, are keen to these dimensions, is too simplistic a proposition, and it carries with it the danger of white scholars ceding expertise to nonwhite scholars (as if, when it comes to race studies, people of colour were the real experts) or of nonwhite scholars absolving themselves of genuinely reflexive practices.” }\citep[p.~582]{emirbayer_race_2012}.\\\textit{“Vast literature explores tensions on this point: on one hand, lived experience undoubtedly shapes one’s understanding of an issue or community, but on the other, one person’s lived experience should neither be enough to speak on behalf of a group [36] nor should it be taken as proof of high-quality work [31].” }\citep[p.~1196]{schroeder_disclosure_2025}.} \\
Theme: Moving beyond individual positionality to consider relationality &  \\
Description  & This theme prompts for considering positionality in more active ways and reflexivity beyond positionality.~  \\
Relevance to structured documentation & Reflexive data work engages critically with this theme by partaking in active reflexivity practices and discussing group or collective reflexivity in addition to self reflexivity.  \\
Examples & {\textit{“By and large, reflexivity often has been conceived in too narrow and underdeveloped a fashion: what the vast majority of thinkers typically have understood as reflexivity has been the exercise of recognizing how aspects of one’s identity or social location can affect one’s vision of the social world. Such a view of reflexivity is necessary but insufficient.” }\citep[p.~577]{emirbayer_race_2012}.\\\textit{“That is, reflexivity is a matter not of plumbing the subjective depths and reconstructing intimate lived experience narrating, for example, one’s own or others’ life-histories but of engaging in rigorous institutional analyses of the social and historical structures that condition one’s thinking and inner experience. Individuals do not come into the world endowed with prenotions; they are the products … of institutions.” }\citep[p.~591]{emirbayer_race_2012}.\\\textit{“First, discussions of epistemic reflexivity often tend to interpret its practice in terms of methodological individualism, which when enacted results in recursive regression and narcissism. Second, this individualist interpretation itself follows from the lack of a collective means for undertaking epistemic reflexivity that is not based on the social field of positions of a field.” }\citep[p.~58]{maton_reflexivity_2003}.\\\textit{“I am particularly concerned with how self-reflexivity may result in a simple identifying of oneself or a telling of a confessional tale, which certainly continues to work to identify and define the “other,” and how the use of self-reflexivity is often used to situate oneself as closer to the subject. This can lead to a specific form of self-reflexivity – a reflexivity that falls into seeking similarities between the researcher and the subject, a reflexivity that seeks to make “your” self closer to “your” subject.”~}\citep[p.~182]{pillow_confession_2003}.}  \\
Theme: Accountability beyond transparency  &  \\
Description  & This theme differentiates between accountability and transparency as features of reflexive research where accountability is taking ownership of the research and its potential negative repercussions while transparency is considered the mere identification of biases and harmful impacts.  \\
Relevance to structured documentation & Structured data documentation is often discussed as a tool for transparency but this alone is not equivalent to reflexivity. Does the structured documentation demonstrate accountability or exclusively and non-reflexively document process?   \\
Examples & {\textit{“For example, again, preregistration of sample size, characteristics of said sample, and recruitment strategies may be a useful~ starting point to embedding transparency into the process (however, it should be reiterated that preregistration does not constitute reflexivity by itself).” }\citep[p.~7]{jamieson_reflexivity_2023}.\\\textit{“Incorporating reflexivity throughout the research process… allows the researcher to constantly be mindful of their actions and roles, and provides an additional layer of scrutiny for their research. Furthermore, Mauthner and Doucet suggested there will be an increase in confidence in research when ``more researchers can be self-conscious about, and articulate, their role in the research process and products.'' }\citep[p.~3]{boyd_quinta_2023}.\\\textit{“A reflexive stance acknowledges that subjectivity and bias are not aberrations that can ever be fully eradicated from research but~ inherent aspects of human inquiry that should be acknowledged and accounted for. As such, we see reflexivity~ as a complement to the push for transparency that is already underway in data science—a complement that is~ necessary to fulfill the potential of data science methods for understanding the social world.” }\citep[p.~13]{tanweer_why_2021}.}  
\end{longtblr}

\section{Additional Quantitative Results from CADA} \label{appendixd}

Complete code and data are in a GitHub repository: https://github.com/zogheibc/structuredreflexivity. 

The extension of Table \ref{tbl:tbl2} is provided below. 

\definecolor{Black}{rgb}{0,0,0}
\begin{longtblr}[
  theme = mytheme,
  label = {tbl:tbl6},
  caption = {The occurrences of keywords related to reflexivity themes in completed datasheets.},
]{
  width = \linewidth,
  colspec = {Q[415]Q[152]Q[330]},
  cell{2}{1} = {r=3}{},
  cell{5}{1} = {r=5}{},
  cell{10}{1} = {r=4}{},
  cell{14}{1} = {r=4}{},
  cell{18}{1} = {r=4}{},
  cell{22}{1} = {r=4}{},
  cell{26}{1} = {r=2}{},
  vlines = {Black},
  vline{1} = {1-2,5,10,14,18,22,26}{black},
  hline{1-2,5,10,14,18,22,26,28} = {-}{},
  hline{3-4,6-9,11-13,15-17,19-21,23-25,27} = {2-3}{},
}
Reflexivity and Themes & Theme Keywords & Number of Occurrences Across Corpus\\
Reflexivity & Reflexivity & 0\\
 & Reflexive & 0\\
 & Reflection & 0\\
Marginalization and power & Marginalization & 0\\
 & Power & 3\\
 & Hierarchy & 1\\
 & Oppression & 0\\
 & Inequity & 0\\
Attention to historical context & Historical & 5\\
 & Social & 34\\
 & Cultural & 1\\
 & Socio-Cultural & 0\\
Attention to scholarly context & Domain & 43\\
 & Discipline & 0\\
 & Institution & 7\\
 & Norms & 0\\
Questioning claims of neutrality & Neutrality & 0\\
 & Objectivity & 0\\
 & Subjectivity & 1\\
 & Bias & 13\\
Moving beyond individual positionality to consider relationality & Positionality & 0\\
 & Identity & 5\\
 & Relationality & 0\\
 & Disclosure & 0\\
Accountability beyond transparency & Accountability & 1\\
 & Transparency & 1
\end{longtblr}

We calculated collocations (bigrams and trigrams) to explore those terms in completed datasets that co-occur most frequently (see Table \ref{tab:tab7} and Table \ref{tab:tab8}). We find that the majority of ngrams are relevant to dataset-specific topics, rather than to themes of reflexivity. 

\begin{table*}[h!]
    \centering
\caption{Top 10 bigrams from completed structured documents.}
\label{tab:tab7}
    \begin{tabular}{|l|l|}\hline
         Bigram& Frequency\\\hline
         chronicling america& 3\\\hline
         hong kong & 3\\\hline
         ego exo& 4\\\hline
         kth royal& 4\\\hline
         raesetje sefala& 4\\\hline
         california irvine& 3\\\hline
         energy consumption& 3\\\hline
         mental workload& 4\\\hline
         mobile phone& 5\\ \hline
 nanjing changzhou&5\\\hline
    \end{tabular}

\end{table*}

\begin{table*}[h!]
    \centering
\caption{Top 10 trigrams from completed structured documents}
\label{tab:tab8}
    \begin{tabular}{|l|l|}\hline
         Trigram & Frequency\\\hline
         ego exo top& 4\\\hline
         export control regulatory& 3\\\hline
         nanjing changzhou wuhan& 5\\\hline
         uv texture map & 4\\\hline
         offensive insulting threatening& 6\\\hline
         view ego exo& 4\\\hline
         experimentally determined structure& 3\\\hline
         flame parameter uv& 3\\\hline
         kth royal institute& 4\\ \hline
 parameter uv texture&3\\\hline
    \end{tabular}

\end{table*}

We also include complete lexical diversity results (the ratio of unique words to total words) for each individual dataset documentation (Table \ref{tab:tab9}), and for each question across all datasheets (Table \ref{tab:tab10}).

\begin{longtable}{|>{\hspace{0pt}}m{0.256\linewidth}|>{\hspace{0pt}}m{0.231\linewidth}|}
\caption{Lexical Diversity per structured document.\label{tab:tab9}}\\ 
\hline
Paper & Lexical Diversity \endfirsthead 
\hline
ambiguous & 0.548589 \\ 
\hline
amerstories & 0.519862 \\ 
\hline
bitod & 0.549550 \\ 
\hline
bubbleML & 0.572816 \\ 
\hline
caesar & 0.406685 \\ 
\hline
cede & 0.495000 \\ 
\hline
change\_event & 0.458977 \\ 
\hline
conflab & 0.411517 \\ 
\hline
covid & 0.516181 \\ 
\hline
cpd & 0.366864 \\ 
\hline
creak & 0.573386 \\ 
\hline
csaw & 0.448760 \\ 
\hline
datacomp & 0.415936 \\ 
\hline
flair & 0.523871 \\ 
\hline
kespeech & 0.625899 \\ 
\hline
loveda & 0.466790 \\ 
\hline
madlad & 0.528037 \\ 
\hline
mathgpt & 0.486264 \\ 
\hline
medsat & 0.463830 \\ 
\hline
objaverse & 0.516432 \\ 
\hline
openprotein & 0.454894 \\ 
\hline
program\_puzzles & 0.560440 \\ 
\hline
prospect & 0.000000 \\ 
\hline
pug & 0.447537 \\ 
\hline
refinedweb & 0.589958 \\ 
\hline
renderme & 0.469363 \\ 
\hline
scamps & 0.513181 \\ 
\hline
scigen & 0.402945 \\ 
\hline
sign\_lang & 0.462500 \\ 
\hline
spatial\_apart & 0.382731 \\ 
\hline
spoken\_lang & 0.466742 \\ 
\hline
star & 0.453532 \\ 
\hline
tufts & 0.480442 \\ 
\hline
visalign & 0.632576 \\ 
\hline
wikichurches & 0.471475 \\
\hline
\end{longtable}

\begin{longtable}{|>{\hspace{0pt}}m{0.150\linewidth}|>{\hspace{0pt}}m{0.150\linewidth}|}
\caption{Lexical diversity across all documents for each question.\label{tab:tab10}}\\ 
\hline
Question~ & Lexical Diversity \endfirsthead 
\hline
1 & 0.473491 \\ 
\hline
2 & 0.572687 \\ 
\hline
3 & 0.567227 \\ 
\hline
4 & 0.781457 \\ 
\hline
5 & 0.540218 \\ 
\hline
6 & 0.492323 \\ 
\hline
7 & 0.559624 \\ 
\hline
8 & 0.483243 \\ 
\hline
9 & 0.591422 \\ 
\hline
10 & 0.627717 \\ 
\hline
11 & 0.606918 \\ 
\hline
12 & 0.459016 \\ 
\hline
13 & 0.589796 \\ 
\hline
14 & 0.563786 \\ 
\hline
15 & 0.581699 \\ 
\hline
16 & 0.634686 \\ 
\hline
17 & 0.576017 \\ 
\hline
18 & 0.586957 \\ 
\hline
19 & 0.605714 \\ 
\hline
20 & 0.941176 \\ 
\hline
21 & 0.617949 \\ 
\hline
22 & 0.568579 \\ 
\hline
23 & 0.709360 \\ 
\hline
24 & 0.564743 \\ 
\hline
25 & 0.567130 \\ 
\hline
26 & 0.619565 \\ 
\hline
27 & 0.564854 \\ 
\hline
28 & 0.643077 \\ 
\hline
29 & 0.643836 \\ 
\hline
30 & 0.615894 \\ 
\hline
31 & 0.688679 \\ 
\hline
32 & 0.827586 \\ 
\hline
33 & 0.588457 \\ 
\hline
34 & 0.570423 \\ 
\hline
35 & 0.578947 \\ 
\hline
36 & 0.926829 \\ 
\hline
37 & 0.511323 \\ 
\hline
38 & 0.623482 \\ 
\hline
39 & 0.532560 \\ 
\hline
40 & 0.692449 \\ 
\hline
41 & 0.673780 \\ 
\hline
42 & 1.000000 \\ 
\hline
43 & 0.546075 \\ 
\hline
44 & 0.457831 \\ 
\hline
45 & 0.508021 \\ 
\hline
46 & 0.464373 \\ 
\hline
47 & 0.757576 \\ 
\hline
48 & 0.571429 \\ 
\hline
49 & 0.875000 \\ 
\hline
50 & 0.606557 \\ 
\hline
51 & 0.572193 \\ 
\hline
52 & 0.584507 \\ 
\hline
53 & 0.553922 \\ 
\hline
54 & 0.698113 \\ 
\hline
55 & 0.533074 \\ 
\hline
56 & 0.544031 \\ 
\hline
57 & 0.900000 \\
\hline
\end{longtable}

\begin{figure*}[h!]
    \centering
    \includegraphics[width=1\linewidth]{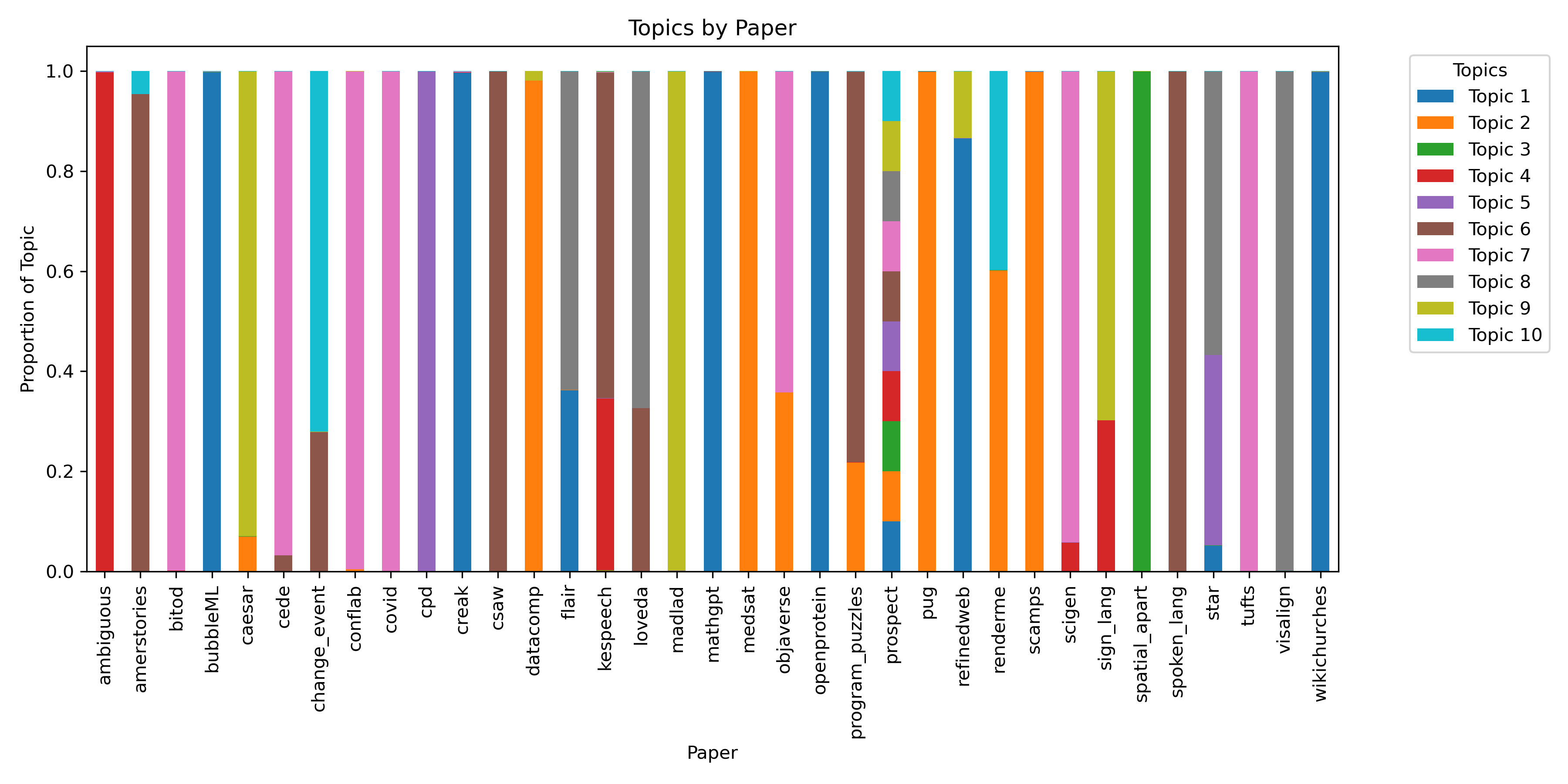}
    \caption{Distribution of topics across individual papers.}
    \label{fig:fig3}
\end{figure*}

\begin{figure*}[h!]
    \centering
    \includegraphics[width=1\linewidth]{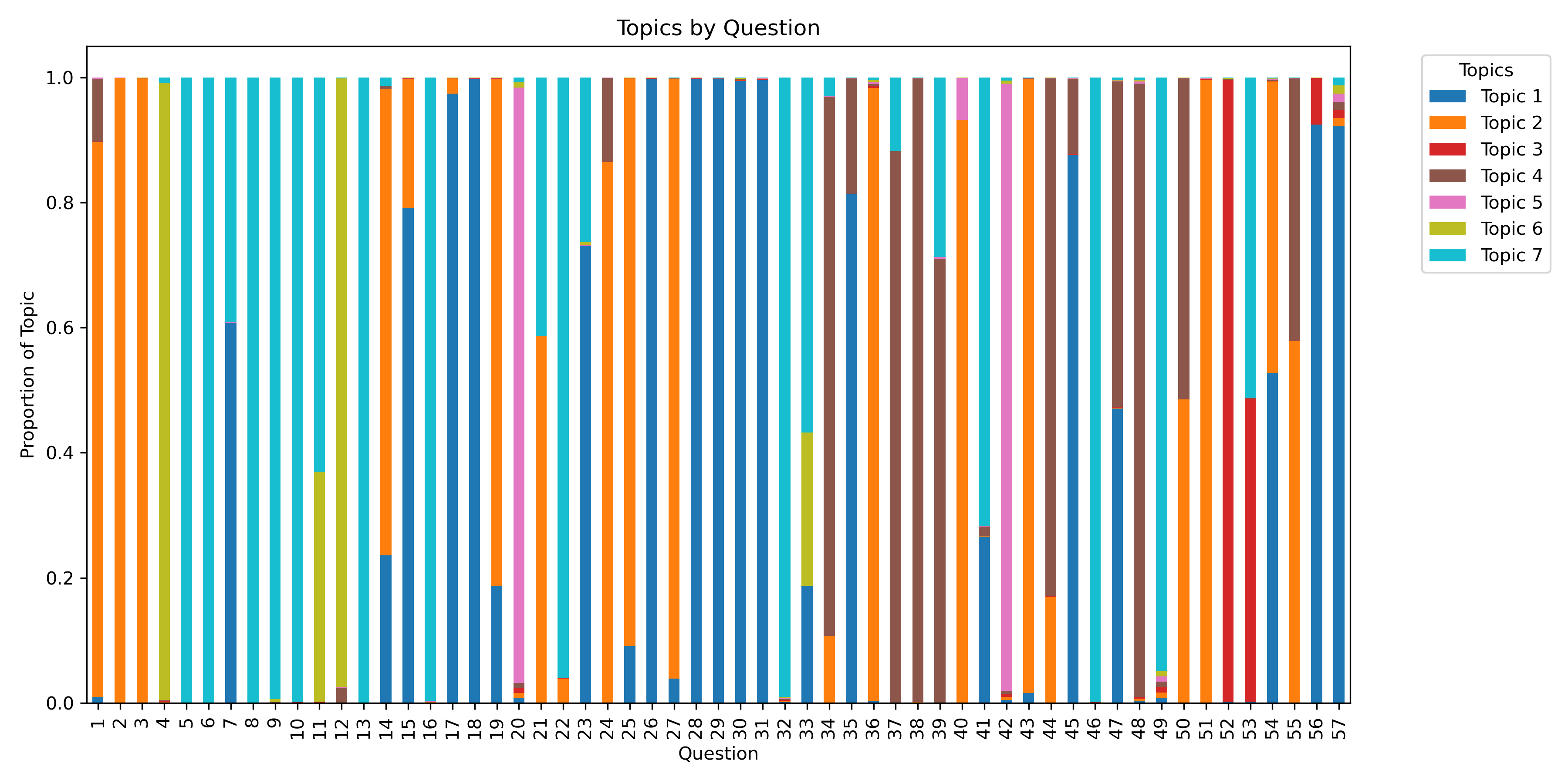}
    \caption{Distribution of topics across individual questions.}
    \label{fig:fig4}
\end{figure*}

We describe in the main text of the paper (Section \ref{findings_quantCADA}) the challenges of topic modeling on this dataset and the implications of these challenges for exploring reflexivity. Here, we include definitions of topics and distributions of topics across papers (Table \ref{tbl:tbl11}, Figure \ref{fig:fig3}) and across questions (Table \ref{tbl:tbl12}, Figure \ref{fig:fig4}). We can observe that topics are relatively more dispersed across individual datasheets, while topics 1 and 7 dominate across many different individual questions. This could imply that dataset creators are providing similar information in their answers to multiple questions on datasheets, limiting deep reflexivity in their answers, but we caution overreliance on these results based on the low coherence and intertopic distances described in the main body of the paper.

\definecolor{Black}{rgb}{0,0,0}
\begin{longtblr}[
  theme = mytheme, 
  label = {tbl:tbl11},
  caption = {Topics and most associated terms generated through LDA topic modeling, treating individual datasheets as documents.},
]{
  width = \linewidth,
  colspec = {Q[60]Q[883]},
  hlines,
  vlines = {Black},
  vline{1} = {-}{black},
}
Topic & Keywords and Probabilities\\
Topic 1 & 0.032*"dataset" + 0.018*"data" + 0.010*"image" + 0.006*"available" + 0.005*"model" + 0.005*"church" + 0.004*"people" + 0.004*"instance" + 0.004*"yes" + 0.004*"see"\\
Topic 2 & 0.029*"data" + 0.027*"dataset" + 0.011*"image" + 0.007*"pug" + 0.006*"available" + 0.005*"used" + 0.005*"instance" + 0.005*"video" + 0.005*"http" + 0.005*"research"\\
Topic 3 & 0.032*"dataset" + 0.023*"data" + 0.009*"image" + 0.006*"used" + 0.005*"yes" + 0.005*"also" + 0.005*"instance" + 0.005*"using" + 0.004*"building" + 0.004*"available"\\
Topic 4 & 0.022*"dataset" + 0.019*"data" + 0.010*"image" + 0.009*"yes" + 0.008*"instance" + 0.006*"video" + 0.006*"paper" + 0.005*"used" + 0.005*"annotation" + 0.005*"available"\\
Topic 5 & 0.027*"data" + 0.021*"dataset" + 0.009*"image" + 0.005*"used" + 0.005*"video" + 0.005*"request" + 0.005*"raw" + 0.005*"officer" + 0.005*"annotation" + 0.004*"available"\\
Topic 6 & 0.030*"dataset" + 0.017*"data" + 0.011*"image" + 0.007*"annotation" + 0.007*"instance" + 0.006*"people" + 0.006*"used" + 0.005*"change" + 0.005*"information" + 0.005*"available"\\
Topic 7 & 0.026*"data" + 0.025*"dataset" + 0.007*"yes" + 0.007*"annotation" + 0.006*"information" + 0.005*"available" + 0.005*"paper" + 0.005*"task" + 0.005*"used" + 0.004*"http"\\
Topic 8 & 0.025*"dataset" + 0.017*"image" + 0.012*"data" + 0.005*"model" + 0.005*"used" + 0.005*"instance" + 0.004*"yes" + 0.004*"people" + 0.004*"task" + 0.004*"annotation"\\
Topic 9 & 0.030*"dataset" + 0.017*"data" + 0.012*"image" + 0.006*"used" + 0.005*"people" + 0.005*"datasets" + 0.004*"language" + 0.004*"yes" + 0.004*"version" + 0.004*"use"\\
Topic 10 & 0.026*"dataset" + 0.013*"data" + 0.011*"image" + 0.006*"change" + 0.006*"annotation" + 0.005*"model" + 0.005*"yes" + 0.005*"used" + 0.005*"event" + 0.005*"information"
\end{longtblr}

\definecolor{Black}{rgb}{0,0,0}
\begin{longtblr}[
  theme = mytheme,
  label = {tbl:tbl12},
  caption = {Topics and most associated terms generated through LDA topic modeling, treating all answers to particular questions as documents.},
]{
  width = \linewidth,
  colspec = {Q[56]Q[888]},
  hlines,
  vlines = {Black},
  vline{1} = {-}{black},
}
Topic & Keywords and Probabilities\\
Topic 1 & 0.035*"dataset" + 0.025*"data" + 0.008*"people" + 0.007*"image" + 0.007*"yes" + 0.006*"available" + 0.005*"individual" + 0.005*"collection" + 0.004*"annotation" + 0.004*"also"\\
Topic 2 & 0.033*"dataset" + 0.024*"data" + 0.010*"image" + 0.007*"available" + 0.006*"information" + 0.005*"research" + 0.005*"yes" + 0.004*"model" + 0.004*"annotation" + 0.004*"http"\\
Topic 3 & 0.025*"dataset" + 0.012*"data" + 0.006*"image" + 0.005*"yes" + 0.005*"people" + 0.004*"github" + 0.004*"used" + 0.004*"annotation" + 0.004*"update" + 0.004*"information"\\
Topic 4 & 0.030*"dataset" + 0.021*"data" + 0.011*"image" + 0.008*"used" + 0.006*"http" + 0.006*"task" + 0.006*"available" + 0.006*"model" + 0.005*"instance" + 0.005*"use"\\
Topic 5 & 0.021*"dataset" + 0.011*"data" + 0.006*"model" + 0.006*"task" + 0.005*"research" + 0.005*"yes" + 0.005*"used" + 0.005*"image" + 0.004*"datasets" + 0.004*"available"\\
Topic 6 & 0.017*"data" + 0.014*"image" + 0.009*"dataset" + 0.007*"instance" + 0.006*"yes" + 0.006*"annotation" + 0.005*"split" + 0.004*"used" + 0.004*"task" + 0.004*"set"\\
Topic 7 & 0.023*"dataset" + 0.022*"data" + 0.014*"image" + 0.008*"instance" + 0.007*"used" + 0.006*"annotation" + 0.005*"license" + 0.005*"yes" + 0.005*"using" + 0.004*"video"
\end{longtblr}

\section{Extended Recommendations} \label{appendixe}

\begin{small}
\definecolor{Black}{rgb}{0,0,0}
\definecolor{Alto}{rgb}{0.85,0.85,0.85}
\begin{longtblr}[
  theme = mytheme,
  label = {tbl:tbl13},
  caption = {Extended recommendations on incorporating reflexivity in datasheet question prompts (extension of Table \ref{tbl:tbl2}). Grey cells indicate questions not relevant to themes of reflexivity. },
]{
  width = \linewidth,
  colspec = {Q[27]Q[333]Q[267]Q[313]},
  cells = {t},
  cell{1}{1} = {c=2}{0.362\linewidth},
  cell{5}{3} = {Alto},
  cell{5}{4} = {Alto},
  cell{6}{3} = {Alto},
  cell{6}{4} = {Alto},
  cell{7}{3} = {Alto},
  cell{7}{4} = {Alto},
  cell{9}{3} = {Alto},
  cell{9}{4} = {Alto},
  cell{12}{3} = {Alto},
  cell{12}{4} = {Alto},
  cell{15}{3} = {Alto},
  cell{15}{4} = {Alto},
  cell{16}{3} = {Alto},
  cell{16}{4} = {Alto},
  cell{17}{3} = {Alto},
  cell{17}{4} = {Alto},
  cell{18}{3} = {Alto},
  cell{18}{4} = {Alto},
  cell{19}{3} = {Alto},
  cell{19}{4} = {Alto},
  cell{20}{3} = {Alto},
  cell{20}{4} = {Alto},
  cell{21}{3} = {Alto},
  cell{21}{4} = {Alto},
  cell{26}{3} = {Alto},
  cell{26}{4} = {Alto},
  cell{28}{3} = {Alto},
  cell{28}{4} = {Alto},
  cell{33}{3} = {Alto},
  cell{33}{4} = {Alto},
  cell{34}{3} = {Alto},
  cell{34}{4} = {Alto},
  cell{35}{3} = {Alto},
  cell{35}{4} = {Alto},
  cell{36}{3} = {Alto},
  cell{36}{4} = {Alto},
  cell{37}{3} = {Alto},
  cell{37}{4} = {Alto},
  cell{38}{3} = {Alto},
  cell{38}{4} = {Alto},
  cell{39}{3} = {Alto},
  cell{39}{4} = {Alto},
  cell{40}{3} = {Alto},
  cell{40}{4} = {Alto},
  cell{43}{3} = {Alto},
  cell{43}{4} = {Alto},
  cell{45}{3} = {Alto},
  cell{45}{4} = {Alto},
  cell{46}{3} = {Alto},
  cell{46}{4} = {Alto},
  cell{47}{3} = {Alto},
  cell{47}{4} = {Alto},
  cell{48}{3} = {Alto},
  cell{48}{4} = {Alto},
  cell{49}{3} = {Alto},
  cell{49}{4} = {Alto},
  cell{50}{3} = {Alto},
  cell{50}{4} = {Alto},
  cell{51}{3} = {Alto},
  cell{51}{4} = {Alto},
  cell{52}{3} = {Alto},
  cell{52}{4} = {Alto},
  cell{54}{3} = {Alto},
  cell{54}{4} = {Alto},
  cell{56}{3} = {Alto},
  cell{56}{4} = {Alto},
  cell{57}{3} = {Alto},
  cell{57}{4} = {Alto},
  cell{58}{3} = {Alto},
  cell{58}{4} = {Alto},
  hlines,
  vlines = {Black},
  vline{1} = {-}{black},
}
Datasheet Questions (from \cite{gebru_datasheets_2021}) &  & Relevant themes (identified in Stage 2) & How to incorporate reflexivity into the question\\
1 & For what purpose was the dataset created? Was there a specific task in mind? Was there a specific gap that needed to be filled? Please provide a description. & {\labelitemi\hspace{\dimexpr\labelsep+0.5\tabcolsep}Attention to scholarly context\\\labelitemi\hspace{\dimexpr\labelsep+0.5\tabcolsep}Questioning claims of neutrality\\\labelitemi\hspace{\dimexpr\labelsep+0.5\tabcolsep}Accountability beyond transparency~} & {Problem formulation is dependent both on the norms of the domain as well as the researchers’ worldviews, expertise, etc. It is a value-laden process composed of several decisions that should be articulated to clarify the context of the problem. The question can prompt for reflexivity by asking for reflection on the process of problem formulation.~\\An extension of the datasheet question could include: \textit{How was this gap or task identified? Was the gap identified by the people affected by this dataset?}}\\
2 & Who created the dataset (e.g., which team, research group) and on behalf of which entity (e.g., company, institution, organization)? & {\labelitemi\hspace{\dimexpr\labelsep+0.5\tabcolsep}Attention to scholarly context\\\labelitemi\hspace{\dimexpr\labelsep+0.5\tabcolsep}Questioning claims of neutrality\\\labelitemi\hspace{\dimexpr\labelsep+0.5\tabcolsep}Moving beyond individual positionality to consider relationality\\\labelitemi\hspace{\dimexpr\labelsep+0.5\tabcolsep}Accountability beyond transparency} & {The dataset is shaped by the researchers and is never a “neutral” or “standard” process. Institutions and organizations can have their own set of scholarly norms and standards that influence how the dataset is developed. Additionally researchers themselves influence the dataset development process. The question can prompt for reflexivity by encouraging authors to provide individual and collective positionality statements.~\\\textit{What roles and hierarchies are present on the research team? Does the entity or organization impose restrictions or limitations on dataset subjects or processes?}}\\
3 & Who funded the creation of the dataset? If there is an associated grant, please provide the name of the grantor and the grant name and number. & {\labelitemi\hspace{\dimexpr\labelsep+0.5\tabcolsep}Marginalization and power~\\\labelitemi\hspace{\dimexpr\labelsep+0.5\tabcolsep}Attention to scholarly context} & {Funding bodies have the ability to influence which topics of research are supported and published. It also demonstrates the norms within the field of what is considered “good” research or deemed research at all.~\\\textit{Are there any limiting factors or conflicts of interest associated with the funding of their dataset?}}\\
4 & Any other comments? &  & \\
5 & What do the instances that comprise the dataset represent (e.g., documents, photos, people, countries)? Are there multiple types of instances (e.g., movies, users, and ratings; people and interactions between them; nodes and edges)? Please provide a description. & The question is about describing dataset contents.~ & \\
6 & How many instances are there in total (of each type, if appropriate)? & The question is about describing dataset contents. & \\
7 & Does the dataset contain all possible instances or is it a sample (not necessarily random) of instances from a larger set? If the dataset is a sample, then what is the larger set? Is the sample representative of the larger set (e.g., geographic coverage)? If so, please describe how this representativeness was validated/verified. If it is not representative of the larger set, please describe why not (e.g., to cover a more diverse range of instances, because instances were withheld or unavailable). & {\labelitemi\hspace{\dimexpr\labelsep+0.5\tabcolsep}Attention to scholarly context\\\labelitemi\hspace{\dimexpr\labelsep+0.5\tabcolsep}Questioning claims of neutrality~\\\labelitemi\hspace{\dimexpr\labelsep+0.5\tabcolsep}Accountability beyond transparency} & {Decisions around sampling and representativeness are embedded in scholarly context and expertise as well as individual experience and training. This question can prompt for reflexivity by asking dataset creators to explain whether different approaches to sampling were considered and what impact these choices may have.\\\textit{What other sampling approaches were considered? Who is relevant but not represented? Why is this tradeoff permissible?~}}\\
8 & What data does each instance consist of? “Raw” data (e.g., unprocessed text or images) or features? In either case, please provide a description. & The question is about describing dataset contents. & \\
9 & Is there a label or target associated with each instance? If so, please provide a description. & \labelitemi\hspace{\dimexpr\labelsep+0.5\tabcolsep}Attention to scholarly context~~ & {The task of associating a label or target to data instances involves converting real-world data into data that can be parsed by a machine learning model. This task involves applying data science knowledge and expertise. The question can prompt for reflexivity by asking dataset creators to explain and reflect on their process of establishing data labels or targets.~\\\textit{How were unclear or conflicted labels addressed? Did data labeling impose binaries or strict categories on indistinct data?}}\\
10 & Is any information missing from individual instances? If so, please provide a description, explaining why this information is missing (e.g., because it was unavailable). This does not include intentionally removed information, but might include, e.g., redacted text. & \labelitemi\hspace{\dimexpr\labelsep+0.5\tabcolsep}Accountability beyond transparency & {This question should prompt dataset creators to consider whether and how the missing information could have an impact on the dataset outcomes.~\\\textit{How would missing data in this dataset impact model results if used for training?~}}\\
11 & Are relationships between individual instances made explicit (e.g., users' movie ratings, social network links)? If so, please describe how these relationships are made explicit. & The question is about describing dataset contents. & \\
12 & Are there recommended data splits (e.g., training, development/validation, testing)? If so, please provide a description of these splits, explaining the rationale behind them. & {\labelitemi\hspace{\dimexpr\labelsep+0.5\tabcolsep}Attention to scholarly context\\\labelitemi\hspace{\dimexpr\labelsep+0.5\tabcolsep}Accountability beyond transparency} & {This question can prompt reflexivity by asking dataset creators to consider how the rationale for the data splits is grounded in disciplinary, domain-specific, or other epistemic values. It can also prompt for reflection on possible impacts.~\\\textit{What impacts would varying data splits have on model outcomes? How may the rationale for the data splits reflect specific epistemic values?}}\\
13 & Are there any errors, sources of noise, or redundancies in the dataset? If so, please provide a description. & \labelitemi\hspace{\dimexpr\labelsep+0.5\tabcolsep}Accountability beyond transparency & {The question can prompt for reflexivity by asking dataset creators to reflect on how errors, noise, redundancies can impact dataset outcomes (beyond providing a description).~\\\textit{What individuals and groups might be impacted by these errors, and in what specific ways? How are these errors being addressed?}}\\
14 & Is the dataset self-contained, or does it link to or otherwise rely on external resources (e.g., websites, tweets, other datasets)? If it links to or relies on external resources, a) are there guarantees that they will exist, and remain constant, over time; b) are there official archival versions of the complete dataset (i.e., including the external resources as they existed at the time the dataset was created); c) are there any restrictions (e.g., licenses, fees) associated with any of the external resources that might apply to a future user? Please provide descriptions of all external resources and any restrictions associated with them, as well as links or other access points, as appropriate. & The question is about describing dataset contents and facilitating reusability. & \\
15 & Does the dataset contain data that might be considered confidential (e.g., data that is protected by legal privilege or by doctor–patient confidentiality, data that includes the content of individuals’ non-public communications)? If so, please provide a description. & This question prompts dataset creators to be \textbf{transparent }about dataset contents. While this question is relevant to themes of marginalization and power, attention to historical context, and accountability beyond transparency, we do not make recommendations to include reflexivity here. The recommendations would prompt for reflexivity regarding consent and impact which are better suited for questions 26,28-31. Thus we recommend datasheet questions 26,28-31 to prompt for reflections on the response provided here.~ & \\
16 & Does the dataset contain data that, if viewed directly, might be offensive, insulting, threatening, or might otherwise cause anxiety? If so, please describe why. & This question prompts dataset creators to be \textbf{transparent }about dataset contents. While this question is relevant to themes of marginalization and power, attention to historical context, and accountability beyond transparency, we do not make recommendations to include reflexivity here. The recommendations would prompt for reflexivity regarding consent and impact which are better suited for questions 26 and 31. Thus we recommend datasheet questions 26 and 31 to prompt for reflections on the response provided here.~ & \\
17 & Does the dataset identify any subpopulations (e.g., by age, gender)? If so, please describe how these subpopulations are identified and provide a description of their respective distributions. & This question prompts dataset creators to be \textbf{transparent }about dataset contents. While this question is relevant to themes of marginalization and power, attention to historical context, and accountability beyond transparency, we do not make recommendations to include reflexivity here. The recommendations would prompt for reflexivity regarding consent and impact which are better suited for questions 26,28-31. Thus we recommend datasheet questions 26,28-31 to prompt for reflections on the response provided here. & \\
18 & Is it possible to identify individuals (i.e., one or more natural persons), either directly or indirectly (i.e., in combination with other data) from the dataset? If so, please describe how. & This question prompts dataset creators to be \textbf{transparent }about dataset contents. While this question is relevant to themes of marginalization and power, attention to historical context, and accountability beyond transparency, we do not make recommendations to include reflexivity here. The recommendations would prompt for reflexivity regarding consent and impact which are better suited for questions 26,28-31. Thus we recommend datasheet questions 26,28-31 to prompt for reflections on the response provided here.~ & \\
19 & Does the dataset contain data that might be considered sensitive in any way (e.g., data that reveals racial or ethnic origins, sexual orientations, religious beliefs, political opinions or union memberships, or locations; financial or health data; biometric or genetic data; forms of government identification, such as social security numbers; criminal history)? If so, please provide a description. & This question prompts dataset creators to be \textbf{transparent }about dataset contents. While this question is relevant to themes of marginalization and power, attention to historical context, and accountability beyond transparency, we do not make recommendations to include reflexivity here. The recommendations would prompt for reflexivity regarding consent and impact which are better suited for questions 26,28-31. Thus we recommend datasheet questions 26,28-31 to prompt for reflections on the response provided here.~ & \\
20 & Any other comments? &  & \\
21 & How was the data associated with each instance acquired? Was the data directly observable (e.g., raw text, movie ratings), reported by subjects (e.g., survey responses), or indirectly inferred/derived from other data (e.g., part-of-speech tags, model-based guesses for age or language)? If data was reported by subjects or indirectly inferred/derived from other data, was the data validated/verified? If so, please describe how. & {\labelitemi\hspace{\dimexpr\labelsep+0.5\tabcolsep}Attention to scholarly context\\\labelitemi\hspace{\dimexpr\labelsep+0.5\tabcolsep}Questioning claims of neutrality\\\labelitemi\hspace{\dimexpr\labelsep+0.5\tabcolsep}Accountability beyond transparency~} & {The decisions relating to data collection sources and methods can reveal assumptions or biases. It can also speak towards nuances of data work, e.g., how data is acquired or derived in cases of poor availability.~\\\textit{How were final data sources chosen? What were the challenges with acquiring the data that influenced the choices?}}\\
22 & What mechanisms or procedures were used to collect the data (e.g., hardware apparatus or sensor, manual human curation, software program, software API)? How were these mechanisms or procedures validated? & {\labelitemi\hspace{\dimexpr\labelsep+0.5\tabcolsep}Attention to scholarly context\\\labelitemi\hspace{\dimexpr\labelsep+0.5\tabcolsep}Accountability beyond transparency~} & {This question can prompt for reflexivity by asking dataset creators to go beyond identifying procedures to also describe why those procedures were adopted and the context surrounding those choices which may reveal nuance and craft around data work. The question can also reinforce discussion on validating the procedures by emphasizing the resulting impact on data authenticity and reliability.~\\\textit{What alternate data collection methods might have been adopted, and why were they not?~}}\\
23 & If the dataset is a sample from a larger set, what was the sampling strategy (e.g., deterministic, probabilistic with specific sampling probabilities)? & {\labelitemi\hspace{\dimexpr\labelsep+0.5\tabcolsep}Attention to scholarly context\\\labelitemi\hspace{\dimexpr\labelsep+0.5\tabcolsep}Questioning claims of neutrality\\\labelitemi\hspace{\dimexpr\labelsep+0.5\tabcolsep}Accountability beyond transparency} & We echo recommendations from datasheet question 7.~\\
24 & Who was involved in the data collection process (e.g., students, crowdworkers, contractors) and how were they compensated (e.g., how much were crowdworkers paid)? & {\labelitemi\hspace{\dimexpr\labelsep+0.5\tabcolsep}Marginalization and power\\\labelitemi\hspace{\dimexpr\labelsep+0.5\tabcolsep}Attention to scholarly context\\\labelitemi\hspace{\dimexpr\labelsep+0.5\tabcolsep}Moving beyond individual positionality to consider relationality} & {In situations where the people involved in data collection are different from those identified in datasheet question 2, there is scope for reflection on how different roles within the knowledge production process may create hierarchies or introduce instances of exploitation and impact the dataset development process. Dataset developers should also reflect on whether their data work could cause harm or violence considering data as akin to bodies, such as through invasive data collection. ~\\The question can prompt for reflexivity on the dynamics within the dataset development group. Dataset creators can also reflect on whether and how experts were recruited for the collection process and how their inclusion impacted the research team’s dynamics.\\\textit{How was fair compensation supported if team members were in different locations with different legal wage standards? Were community members or members of population groups represented in the dataset involved in the dataset collection process? Why or why not? Was it possible for the individuals being researched to be included in the knowledge production process?}}\\
25 & Over what timeframe was the data collected? Does this timeframe match the creation timeframe of the data associated with the instances (e.g., recent crawl of old news articles)? If not, please describe the timeframe in which the data associated with the instances was created. & The question is about describing dataset contents. & \\
26 & Were any ethical review processes conducted (e.g., by an institutional review board)? If so, please provide a description of these review processes, including the outcomes, as well as a link or other access point to any supporting documentation. & {\labelitemi\hspace{\dimexpr\labelsep+0.5\tabcolsep}Marginalization and power\\\labelitemi\hspace{\dimexpr\labelsep+0.5\tabcolsep}Attention to historical context\\\labelitemi\hspace{\dimexpr\labelsep+0.5\tabcolsep}Accountability beyond transparency} & {The datasheet question currently prompts dataset creators to be transparent about ethical review processes and outcomes. Additional reflexivity can be encouraged by asking creators to reflect specifically on their responses for questions 15-19 and consider the proportionality principle which requires weighing the positive impacts of the dataset in comparison to the negative impacts, such as if confidential (question 15) or sensitive data (question 19) is included, if data is offensive or threatening (question 16), and if data identifies subpopulations (question 17) or individuals (question 18) resulting in unfair or unethical treatment.\\\textit{What ethical concerns are important but may not be captured by these institutional processes?~}\\\textit{What standards or reviews would be applied if those offended but not involved had a say?}~}\\
27 & Did you collect the data from the individuals in question directly, or obtain it via third parties or other sources (e.g., websites)? & The question is about describing dataset contents. & \\
28 & Were the individuals in question notified about the data collection? If so, please describe (or show with screenshots or other information) how notice was provided, and provide a link or other access point to, or otherwise reproduce, the exact language of the notification itself. & {\labelitemi\hspace{\dimexpr\labelsep+0.5\tabcolsep}Marginalization and power~\\\labelitemi\hspace{\dimexpr\labelsep+0.5\tabcolsep}Accountability beyond transparency} & {In addition to being transparent about whether and how notification about data collection was provided, the question can prompt dataset creators to consider the roles of the researcher(s) versus the “researched” for their dataset. This reflection should also be made in context to the responses for questions 15, 17-19 (i.e., notification is increasingly important if the dataset has confidential, sensitive, or identifiable content). Responses to this question should additionally consider how marginalized populations may feel targeted for data collection, and how this may continue historically exploitative, oppressive, violent, or extractive practices. \\\textit{If participants were not notified, what was the reason and how does that speak to power differentials between researcher(s) and the “researched”?}}\\
29 & Did the individuals in question consent to the collection and use of their data? If so, please describe (or show with screenshots or other information) how consent was requested and provided, and provide a link or other access point to, or otherwise reproduce, the exact language to which the individuals consented. & {\labelitemi\hspace{\dimexpr\labelsep+0.5\tabcolsep}Marginalization and power~\\\labelitemi\hspace{\dimexpr\labelsep+0.5\tabcolsep}Accountability beyond transparency} & {This question can prompt reflexivity by asking dataset creators to consider the power differentials between the researcher(s) and the “researched” especially in cases where consent was not obtained or is considered implied.~ This reflection should also be made in context to the responses for questions 15, 17-19 (i.e., \textit{informed }consent is increasingly important if the dataset has confidential, sensitive, or identifiable content).\\\textit{Did individuals consent to all potential future uses of their data? Was consent implied (e.g. by website terms and conditions) or informed (e.g. sought out for inclusion in a particular dataset)?}}\\
30 & If consent was obtained, were the consenting individuals provided with a mechanism to revoke their consent in the future or for certain uses? If so, please provide a description, as well as a link or other access point to the mechanism (if appropriate). & {\labelitemi\hspace{\dimexpr\labelsep+0.5\tabcolsep}Marginalization and power~\\\labelitemi\hspace{\dimexpr\labelsep+0.5\tabcolsep}Accountability beyond transparency} & {This question can prompt reflexivity by asking dataset creators to consider the power differentials between the researcher(s) and the “researched” especially in cases where consent cannot be revoked. This reflection should also be made in context to the responses for questions 15, 17-19 (i.e., the ability to revoke consent is increasingly important if the dataset has confidential, sensitive, or identifiable content).\\\textit{How might power dynamics between dataset developers and consenting individuals affect willingness to revoke consent? What steps are being taken to ensure that individuals feel comfortable revoking consent?}}\\
31 & Has an analysis of the potential impact of the dataset and its use on data subjects (e.g., a data protection impact analysis) been conducted? If so, please provide a description of this analysis, including the outcomes, as well as a link or other access point to any supporting documentation. & {\labelitemi\hspace{\dimexpr\labelsep+0.5\tabcolsep}Marginalization and power\\\labelitemi\hspace{\dimexpr\labelsep+0.5\tabcolsep}Attention to scholarly context\\\labelitemi\hspace{\dimexpr\labelsep+0.5\tabcolsep}Accountability beyond transparency} & {The question can further prompt reflexivity by asking dataset creators to consider and justify the chosen method for impact analysis as this choice may reflect epistemological norms or standards.~\\\textit{How does the chosen method for impact analysis reflect epistemological norms or standards? What potential impacts might not be captured by this analysis? What steps were taken to ensure that impacts on individuals and populations were considered in a non-tokenistic manner?}}\\
32 & Any other comments? &  & \\
33 & Was any preprocessing/cleaning/labeling of the data done (e.g., discretization or bucketing, tokenization, part-of-speech tagging, SIFT feature extraction, removal of instances, processing of missing values)? If so, please provide a description. If not, you may skip the remainder of the questions in this section. & This question prompts dataset creators to be \textbf{transparent }about their dataset development process. While this question is relevant to themes of ‘attention to scholarly context’ and ‘accountability beyond transparency’, we do not make recommendations to include reflexivity here. The recommendations would prompt for reflexivity regarding potential impact which is already captured in question 40. & \\
34 & Was the “raw” data saved in addition to the preprocessed/cleaned/labeled data (e.g., to support unanticipated future uses)? If so, please provide a link or other access point to the “raw” data. & The question is about facilitating reusability. & \\
35 & Is the software used to preprocess/clean/label the instances available? If so, please provide a link or other access point. & The question is about facilitating reusability. & \\
36 & Any other comments? &  & \\
37 & Has the dataset been used for any tasks already? If so, please provide a description. & This question prompts dataset creators to be \textbf{transparent }about how their dataset has been used. While this question is relevant to themes of ‘accountability beyond transparency’, we do not make recommendations to include reflexivity here. The recommendations would prompt for reflexivity regarding potential negative uses which is already captured in question 41. & \\
38 & Is there a repository that links to any or all papers or systems that use the dataset? If so, please provide a link or other access point. & The question is about facilitating reusability. & \\
39 & What (other) tasks could the dataset be used for? & This question prompts dataset creators to be \textbf{transparent }about how their dataset can be used. While this question is relevant to themes of ‘accountability beyond transparency’, we do not make recommendations to include reflexivity here. The recommendations would prompt for reflexivity regarding potential negative uses which is already captured in question 41. & \\
40 & Is there anything about the composition of the dataset or the way it was collected and preprocessed/cleaned/labeled that might impact future uses? For example, is there anything that a future user might need to know to avoid uses that could result in unfair treatment of individuals or groups (e.g., stereotyping, quality of service issues) or other undesirable harms (e.g., financial harms, legal risks) If so, please provide a description. Is there anything a future user could do to mitigate these undesirable harms? & {\labelitemi\hspace{\dimexpr\labelsep+0.5\tabcolsep}Attention to scholarly context~\\\labelitemi\hspace{\dimexpr\labelsep+0.5\tabcolsep}Accountability beyond transparency~} & {Data pre-processing involves a series of decisions and choices which reveal values and epistemological norms. This question can prompt dataset creators to be reflexive about the impacts of these choices and make explicit the scholarly context that influences processing decisions.~\\This theme already prompts for reflexivity by encouraging dataset creators to be accountable for undesirable harms by reflecting on mitigation strategies.~\\\textit{What implicit domain knowledge and data skills were applied in the preprocessing of the dataset?~}}\\
41 & Are there tasks for which the dataset should not be used? If so, please provide a description. & \labelitemi\hspace{\dimexpr\labelsep+0.5\tabcolsep}Accountability beyond transparency~ & {The datasheet question currently prompts dataset creators to be transparent about unwanted or harmful uses of the dataset. Further reflexivity can be encouraged by asking creators to consider specifically how such negative uses can be mitigated (similar to question 40).\\\textit{What steps are being taken to prevent misuse of the dataset?}}\\
42 & Any other comments? &  & \\
43 & Will the dataset be distributed to third parties outside of the entity (for example, company, institution, organization) on behalf of which the dataset was created? If so, please provide a description. & \labelitemi\hspace{\dimexpr\labelsep+0.5\tabcolsep}Accountability beyond transparency~ & {The question can prompt dataset creators to reflect whether third parties would take the same considerations towards accountability and transparency as documented here and to what extent the creators of the dataset are responsible for future uses and applications of the dataset.~\\\textit{How will these third parties be held accountable for future usage of the dataset?}}\\
44 & How will the dataset be distributed (e.g., tarball on website, API, GitHub)? Does the dataset have a digital object identifier (DOI)? & The question is about facilitating reusability. & \\
45 & When will the dataset be distributed? & The question is about facilitating reusability. & \\
46 & Will the dataset be distributed under a copyright or other intellectual property (IP) license, and/or under applicable terms of use (ToU)? If so, please describe this license and/or ToU, and provide a link or other access point to, or otherwise reproduce, any relevant licensing terms or ToU, as well as any fees associated with these restrictions. & The question is about facilitating reusability. & \\
47 & Have any third parties imposed IP-based or other restrictions on the data associated with the instances? If so, please describe these restrictions, and provide a link or other access point to, or otherwise reproduce, any relevant licensing terms, as well as any fees associated with these restrictions. & The question is about facilitating reusability. & \\
48 & Do any export controls or other regulatory restrictions apply to the dataset or to individual instances? If so, please describe these restrictions, and provide a link or other access point to, or otherwise reproduce, any supporting documentation. & The question is about facilitating reusability. & \\
49 & Any other comments? &  & \\
50 & Who will be supporting/hosting/maintaining the dataset? & The question is about facilitating reusability. & \\
51 & How can the owner/curator/ manager of the dataset be contacted (for example, email address)? & The question is about facilitating reusability. & \\
52 & Is there an erratum? If so, please provide a link or other access point. & \labelitemi\hspace{\dimexpr\labelsep+0.5\tabcolsep}Accountability beyond transparency~ & {The question can prompt dataset creators to be reflexive particularly if there is no erratum by considering potential downstream impacts if an error is discovered through dataset reuse but not reported/documented.\\\textit{If not, what errors in your dataset and dataset development process might have resulted in damaging consequences?~}}\\
53 & Will the dataset be updated (for example, to correct labeling errors, add new instances, delete instances)? If so, please describe how often, by whom, and how updates will be communicated to dataset consumers (for example, mailing list, GitHub)? & The question is about describing dataset contents and facilitating reusability. & \\
54 & If the dataset relates to people, are there applicable limits on the retention of the data associated with the instances (for example, were the individuals in question told that their data would be retained for a fixed period of time and then deleted)? If so, please describe these limits and explain how they will be enforced. & \labelitemi\hspace{\dimexpr\labelsep+0.5\tabcolsep}Accountability beyond transparency~ & {The question can prompt reflexivity by asking dataset creators to justify the rationale behind the limits on retention. This is aimed to help researchers consider and reflect on whether the retention policy is longer than absolutely needed.~\\\textit{Why are retention limits in place (or not)? What are the environmental or privacy implications of long-term retention?}}\\
55 & Will older versions of the dataset continue to be supported/hosted/ maintained? If so, please describe how. If not, please describe how its obsolescence will be communicated to dataset consumers. & The question is about describing dataset contents and facilitating reusability. & \\
56 & If others want to extend/augment/build on/contribute to the dataset, is there a mechanism for them to do so? If so, please provide a description. Will these contributions be validated/verified? If so, please describe how. If not, why not? Is there a process for communicating/distributing these contributions to dataset consumers? If so, please provide a description. & The question is about describing dataset contents and facilitating reusability. & \\
57 & Any other comments? &  & 
\end{longtblr}
\end{small}

\end{appendix}

\end{document}